\documentclass[pdflatex,sn-mathphys-num,iicol]{sn-jnl}

\usepackage{graphicx}%
\usepackage{subcaption}%
\usepackage{multirow}%
\usepackage{amsmath,amssymb,amsfonts}%
\usepackage{bm}%
\usepackage{amsthm}%
\usepackage{mathrsfs}%
\usepackage[title]{appendix}%
\usepackage{xcolor}%
\usepackage{textcomp}%
\usepackage{manyfoot}%
\usepackage{booktabs}%
\usepackage{algorithm}%
\usepackage{algorithmicx}%
\usepackage{algpseudocode}%
\usepackage{listings}%

\theoremstyle{thmstyleone}%
\theoremstyle{thmstyletwo}%

\theoremstyle{thmstylethree}%

\begin{document}

\title[Structured GSW for GPD-based polarimetry]{Structured generalized sliced Wasserstein distance for keV X-ray polarization analysis with gas pixel detector}


\author*[1,2]{\fnm{Pengcheng} \sur{Ai}}\email{aipc@ccnu.edu.cn}

\author[1,2]{\fnm{Hongtao} \sur{Qin}}

\author[1,2]{\fnm{Xiangming} \sur{Sun}}

\author[1,2]{\fnm{Dong} \sur{Wang}}

\author[3]{\fnm{Huanbo} \sur{Feng}}

\author[3]{\fnm{Hongbang} \sur{Liu}}

\affil*[1]{\orgdiv{PLAC, Key Laboratory of Quark and Lepton Physics (MOE)}, \orgname{Central China Normal University}, \orgaddress{\street{No. 152 Luoyu Road}, \city{Wuhan}, \postcode{430079}, \state{Hubei}, \country{China}}}

\affil[2]{\orgname{Hubei Provincial Engineering Research Center of Silicon Pixel Chip \& Detection Technology}, \orgaddress{\street{No. 152 Luoyu Road}, \city{Wuhan}, \postcode{430079}, \state{Hubei}, \country{China}}}

\affil[3]{\orgdiv{Guangxi Key Laboratory for Relativistic Astrophysics, School of Physical Science and Technology}, \orgname{Guangxi University}, \orgaddress{\city{Nanning}, \postcode{530004}, \country{China}}}


\abstract{Because of the special angular distribution of excited electrons by the photoelectric effect, the Gas Pixel Detector (GPD) is effective in measuring keV X-ray polarization of astrophysical events (e.g. gamma-ray bursts), by capturing ionization tracks of excited electrons as polarized images. Traditionally, the emission angles of photoelectrons are extracted from polarized images first, and statistics are then performed on these angles to infer the polarization direction and intensity. However, observation with the wide field of view requires the incident angle of X-rays not directly attainable through the traditional analysis process. In this paper, we propose using the generalized sliced Wasserstein (GSW) distance, projected by neural networks with random weights, as a completely data-driven approach to analyze X-ray polarization based on two-dimensional polarized images. We find the structures of the randomized neural networks matter when focusing on different aspects of the polarized images, and take advantage of the discrimination abilities by different neural network structures. The proposed method, named the structured GSW distance, successfully distinguishes polarized images with different configurations of incident angles and polarization directions. Furthermore, we build a simplified statistical model based on the von Mises distribution and the circular Wasserstein distance and compare the model against the proposed method, showing their high consistency. The computational method reported in this paper may benefit GPD-based polarimetry in astroparticle experiments and also pattern analysis on raw data from pixel detectors.}




\maketitle

\section{Introduction}
\label{sec:intro}

Pixelated detectors have been coupled with ionizable materials because of their distinct advantage of being able to capture spatially extended ionization tracks. This principle has been adopted in experiments like searching for neutrinoless double-beta decay \cite{Wang_2024,Filipenko2013} and dark matter \cite{instruments6010006}, validating the Migdal effect \cite{Yi2026}, beam monitoring and localization \cite{Wang2022}, and measuring X-ray polarization in astroparticle researches \cite{WEISSKOPF20161179,Feng2019,Feng2024}. The recorded two-dimensional (2D) images have complicated patterns related to rich physical information, making them good candidates for applications of machine learning and neural network techniques \cite{Ai_2018,Amaro2025,AI2020164640,Jiao2025}.

In astrophysics, observing gamma-ray bursts (GRBs) provides insights into the formation and evolution of large celestial bodies. Particularly, measuring the polarization of the soft X-ray \cite{Detlefs2012} (at keV scale) afterglow is an important probe to GRB properties. Gas Pixel Detectors (GPDs), owing to their superior sensitivity to soft X-rays, were used in commissioned spaceborne experiments (IXPE \cite{WEISSKOPF20161179}, PolarLight \cite{Feng2019}, etc.), and will be used in POLAR-2 \cite{Huang2021,Feng2024} in the future. In a GPD, the incident X-ray interacts with the working gas through the photoelectric effect, the primary electron ionizes the gas molecules along its track, and ionization charges drift to the charge multiplier under the electric field before detected by the pixel detector. Traditional data analysis involves reconstructing and counting the emission angles of the photoelectrons in individual signal events to match the non-isotropic angular distribution of polarized X-rays.

Although intuitive and well-explainable, the above two-stage process (extracting emission angles and inferring polarization) relies on elaborately tuning parameters of the handcrafted algorithm \cite{Puetter:1999qb} by human experts, and may gradually lose precision in intermediate steps. Besides, for observation with the wide field of view, analysis of the polarization direction and intensity is entangled with analysis of the incident angle of X-rays, which hinders precise measurement when the image quality is limited. Hence, it would be desirable to devise other ways to utilize the data distribution directly and to fully exploit the physical information lying in the polarized images.

Wasserstein distance \cite{Panaretos2019} is a distance (metric) of measuring the dis-similarity between two data distributions while satisfying four axioms (non-negativity, identity of indiscernibles, symmetry and triangle inequality). The sliced Wasserstein (SW) distance \cite{8578459} and its generalized variant (GSW) \cite{DBLP:conf/nips/KolouriNSBR19} share the statistical properties of the original Wasserstein distance and exhibit as powerful tools to model data distributions in high-dimensional spaces \cite{DBLP:conf/nips/NietertGSK22,8578465}. In the GSW distance, a group of neural networks with randomly initialized weights project the 2D images from the data space to the feature space where the Wasserstein distance is computed.

In this paper, we propose to analyze X-ray polarization based on 2D polarized images using the GSW distance as a completely data-driven approach. An experimental system with the GPD is constructed to collect data from several configurations of incident angles and polarization directions. Convolutional neural networks with dual-branch structures are empirically designed to effectively distinguish between data distributions and to improve the discrimination power. We name the proposed method the structured generalized sliced Wasserstein (SGSW) distance. It is worth mentioning that the method proposed here is different from conventional deep learning approaches based on supervised learning \cite{KITAGUCHI2019162389,Li2025,DBLP:journals/corr/abs-2005-08126}, and also different from neural network-enhanced unsupervised/self-supervised approaches relying on optimizing an efficient feature extractor \cite{Dort2022,DBLP:conf/nips/ChenKSNH20,DBLP:journals/tmlr/OquabDMVSKFHMEA24}. No training is conducted in the process, whether in the supervised way or in the unsupervised way.

The rest of the paper is structured as follows: a prototype of the low-energy polarization detector (LPD) at POLAR-2 is introduced in Sect.~\ref{sec:proto}, with a brief description of the data collection for subsequent analysis. The methodology based on the SGSW distance is explained in detail in Sect.~\ref{sec:method}. The experimental results are given in Sect.~\ref{sec:res}. A simplified statistical model to evaluate the validity of the results is described in Sect.~\ref{sec:stat}. Finally, Sect.~\ref{sec:con} concludes the study.

\section{Prototype of the POLAR-2/LPD}
\label{sec:proto}

\subsection{The detector}

\begin{figure*}[htbp]
	\centering
	\includegraphics[width=0.9\textwidth]{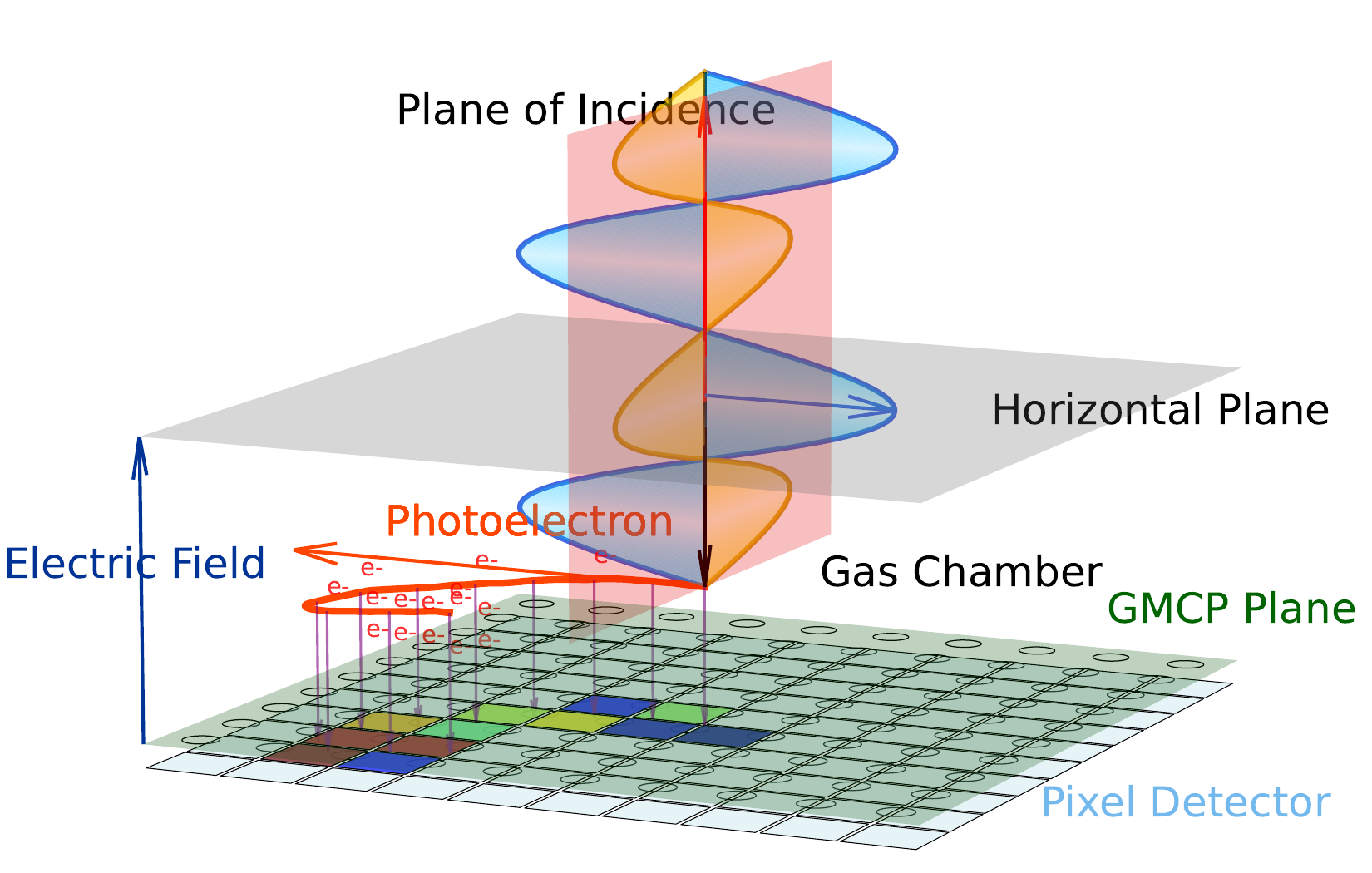}
	\caption{The functional diagram of the GPD based on the gas microchannel plate (GMCP) and the Topmetal pixel detector.}\label{fig:func}
\end{figure*}

\begin{figure*}[htbp]
	\centering
	\begin{subfigure}{0.32\textwidth}
		\includegraphics[width=1.05\textwidth]{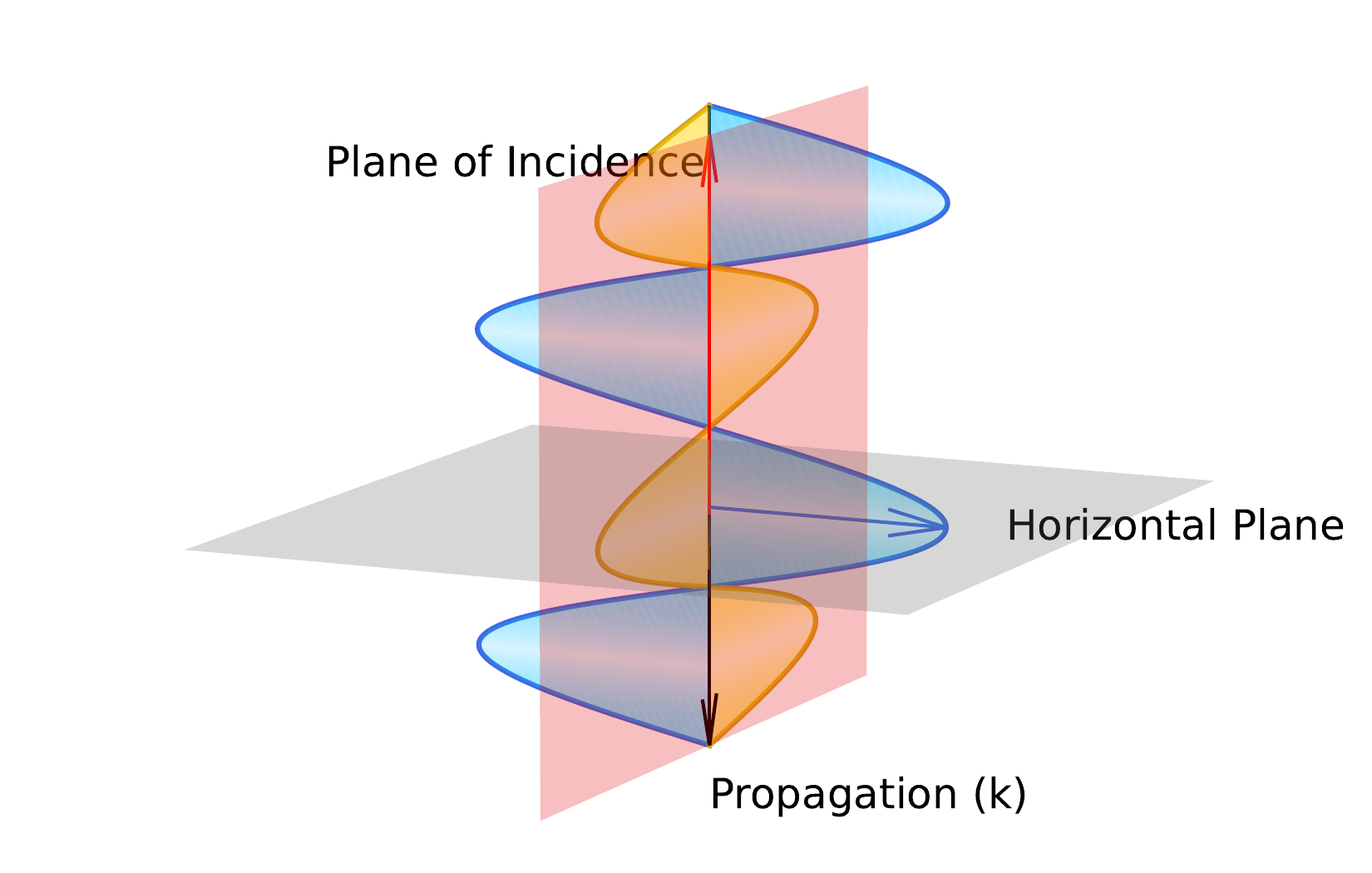}
		\caption{Normal incidence.\\ \emph{(normal\_inc)}}
		\label{fig:norm}
	\end{subfigure}
	\hfill
	\begin{subfigure}{0.32\textwidth}
		\includegraphics[width=1.05\textwidth]{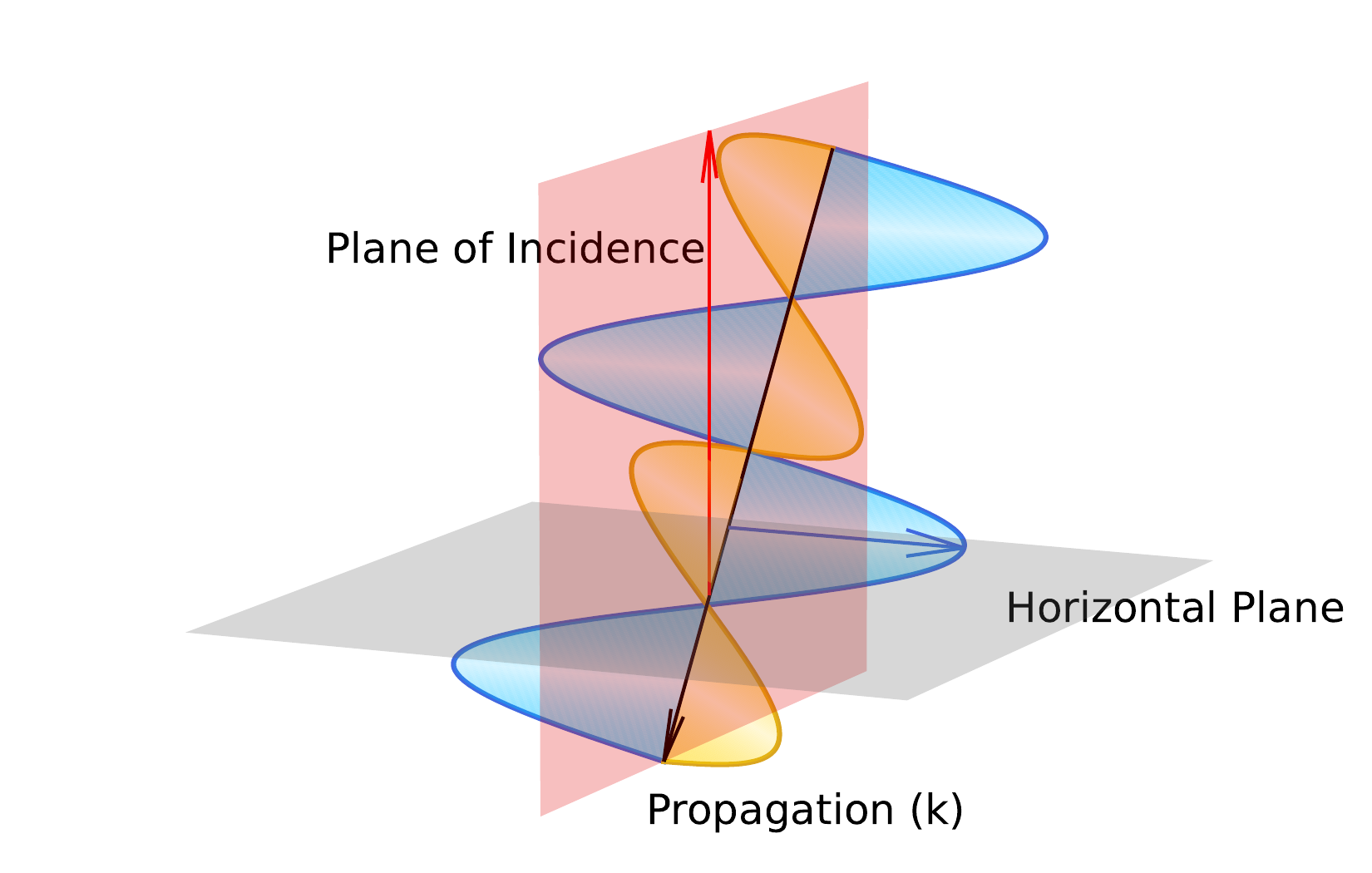}
		\caption{Oblique incidence with the electric vector vertical. \emph{(oblique\_inc\_v)}}
		\label{fig:ov}
	\end{subfigure}
	\hfill
	\begin{subfigure}{0.32\textwidth}
		\includegraphics[width=1.05\textwidth]{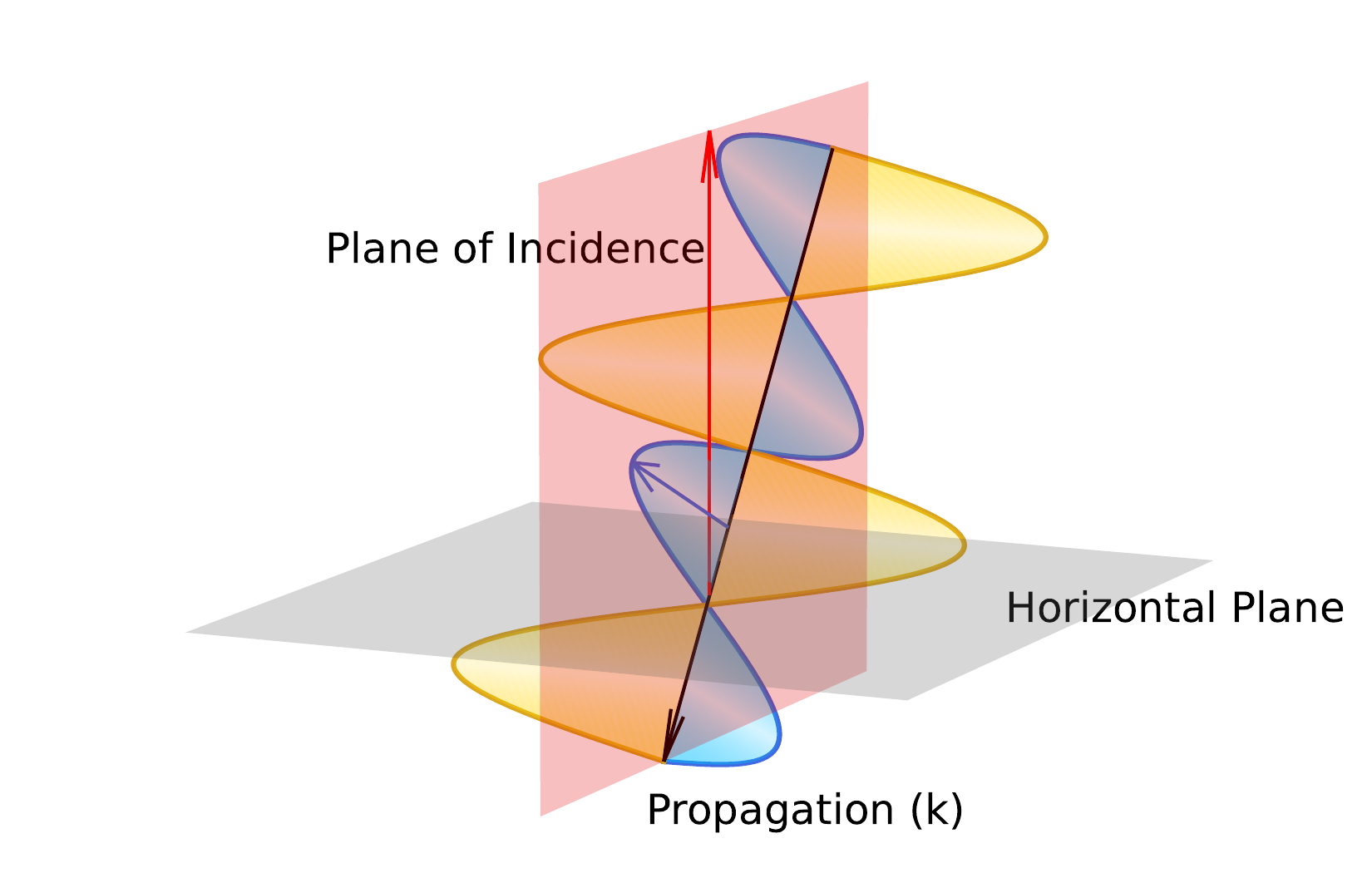}
		\caption{Oblique incidence with the electric vector parallel. \emph{(oblique\_inc\_p)}}
		\label{fig:op}
	\end{subfigure}
	\caption{Illustrations of linearly polarized X-rays with different angles of incidences and different directions of electric vectors. The blue shade indicates the electric vector, and the orange shade indicates the magnetic vector.}
	\label{fig:dir}
\end{figure*}

\begin{figure*}[htbp]
	\centering
	\begin{subfigure}{0.9\textwidth}
		\includegraphics[width=\textwidth]{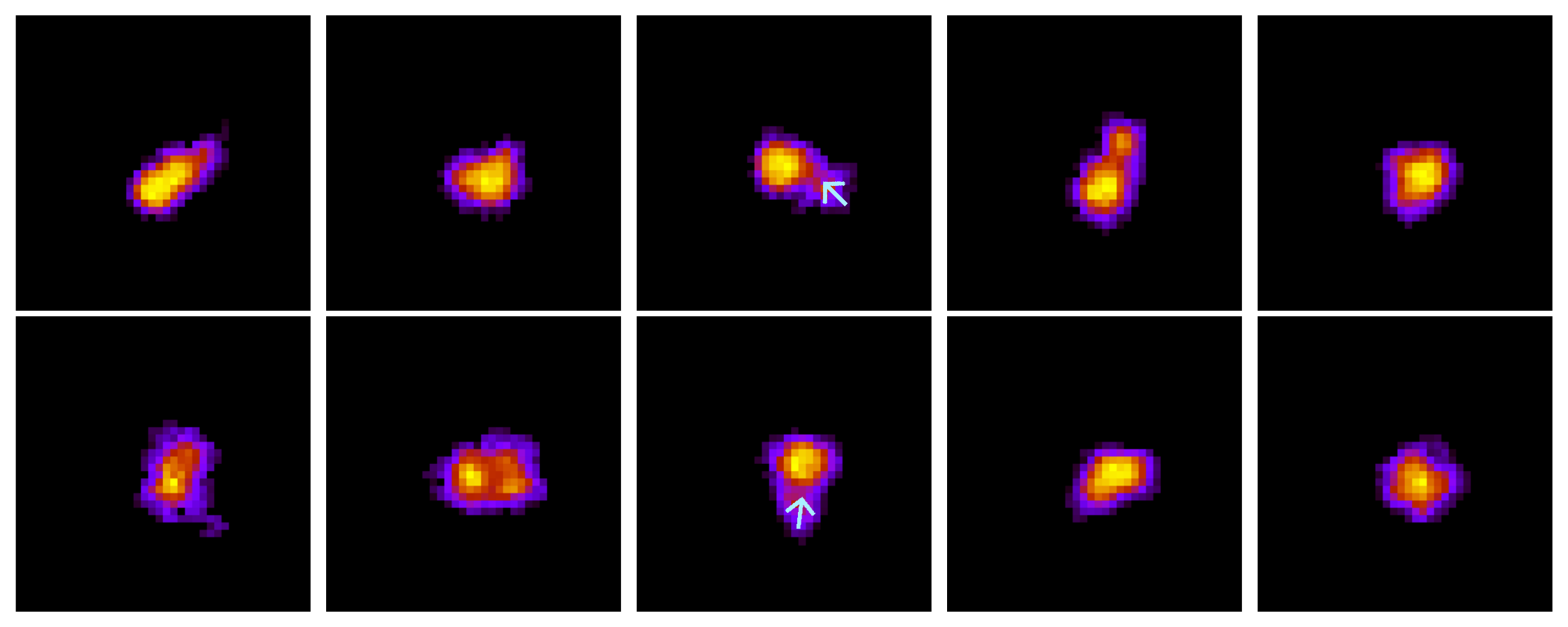}
		\caption{Normal incidence.}
		\label{fig:norm-data}
	\end{subfigure}
	\hfill
	\begin{subfigure}{0.9\textwidth}
		\includegraphics[width=\textwidth]{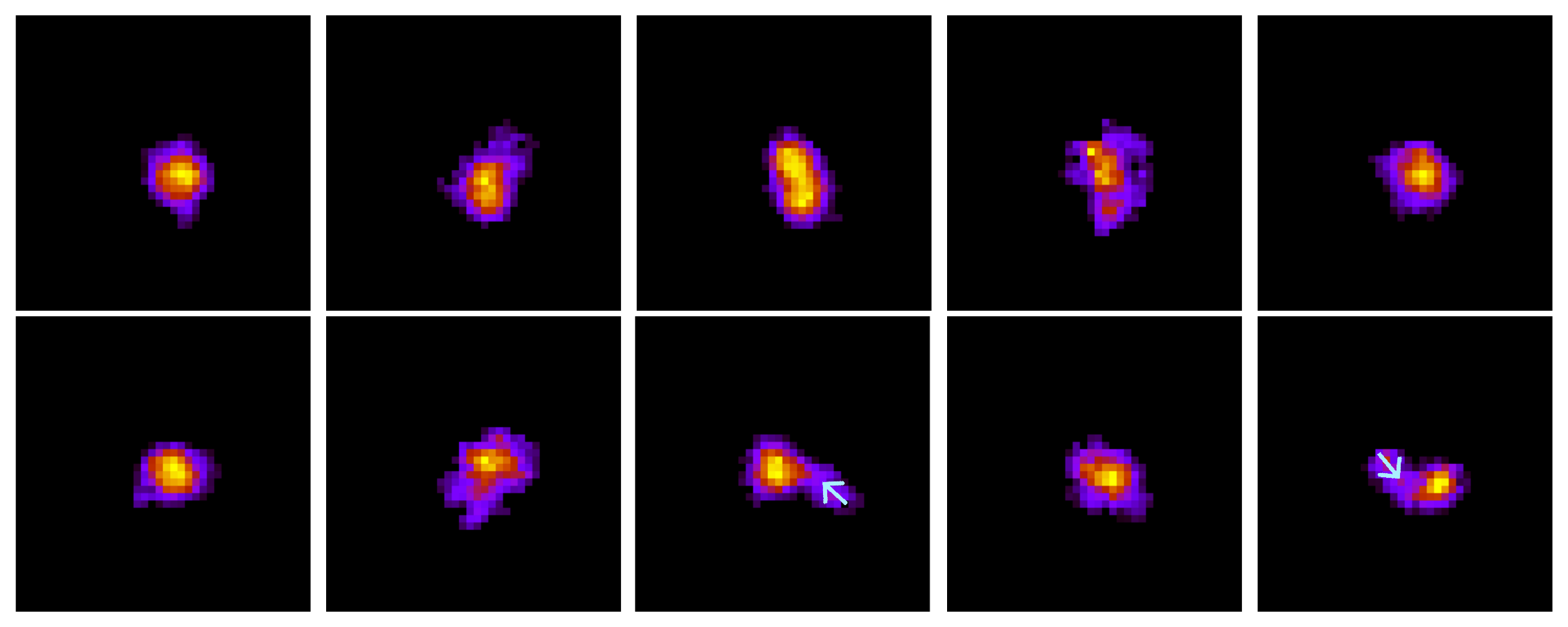}
		\caption{Oblique incidence with the electric vector vertical.}
		\label{fig:ov-data}
	\end{subfigure}
	\hfill
	\begin{subfigure}{0.9\textwidth}
		\includegraphics[width=\textwidth]{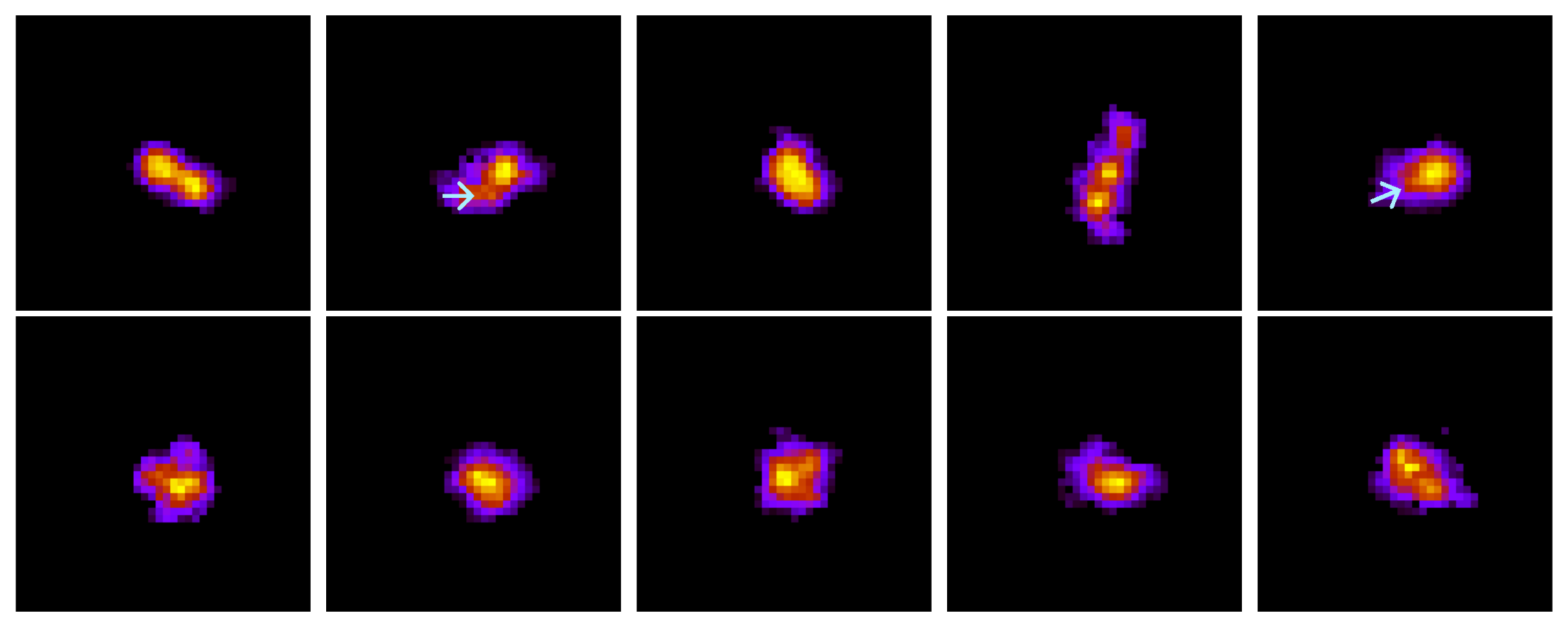}
		\caption{Oblique incidence with the electric vector parallel.}
		\label{fig:op-data}
	\end{subfigure}
	\caption{Illustrations of polarized images with different configurations of incident angles and polarization directions. We crop the 40$\times$40 area in the center of the pixel detector. For each configuration, we annotate two images with the presumed emission direction of the photoelectron.}
	\label{fig:data}
\end{figure*}

POLAR-2 \cite{10.1117/12.2559374} is an international mission with the main goal of observing polarization of GRBs. The LPD \cite{Huang2021,Feng2024} of POLAR-2 is individually developed focusing on the energy range of soft X-rays (2-10 keV). The main apparatus in LPD is a GPD, with a functional diagram shown in Fig.~\ref{fig:func}, based on the gas microchannel plate (GMCP) and the Topmetal \cite{AN2016144,LI2021165430} pixel detector composed of a 72$\times$72 square pixel array (with approx. 83 $\mu$m pixel pitch). The X-rays from GRB sources, with the assumed linear polarization, enter the gas chamber and interact with the working gas (mixture of 40\% helium and 60\% dimethyl ether) through the photoelectric effect. A photoelectron is excited with high probability along the direction of the electric vector. Its angular distribution (K-shell, non-relativistic region) is given by \cite{PhysRev.113.514}:
\begin{equation} \label{equ:angle}
	\frac{\mathrm{d}\sigma_{\mathrm{ph}}}{\mathrm{d}\Omega} \propto \frac{\sin^2 \theta \cos^2 \phi}{(1 - \beta \cos \theta)^4}
\end{equation}

\noindent where $\theta$ is the polar angle relative to the propagation direction, $\phi$ is the azimuthal angle of the emitted photoelectron relative to the electric vector, and $\beta$ is the ratio of the electron speed to the speed of light.

The photoelectron ionizes along its track and forms a Bragg peak at the far end. The ionization charges (electrons) drift under the electric field and multiply at the GMCP. The Topmetal pixel detector is a direct charge sensor without gas-electron multiplication. The ionization charges are recorded by Topmetal and read out by front-end electronics.

A distinct advantage of POLAR-2/LPD is its wide field of view. Incident X-rays are not limited to a narrow collimator, which improves the detection efficiency but also brings about complexities in data analysis. In Fig.~\ref{fig:dir}, three typical configurations of polarization are illustrated and considered for the rest of the paper: (1) \emph{normal\_inc}: the X-ray enters the detector with the propagation direction being normal to the horizontal plane; (2) \emph{oblique\_inc\_v}: the X-ray enters the detector with oblique incidence, and the electric vector is vertical\footnote{We use the term ``vertical'' for X-rays with the electric vector perpendicular to the plane of incidence to avoid confusion with ``parallel'' in the abbreviation.} (perpendicular) to the plane of incidence; and (3) \emph{oblique\_inc\_p}: the X-ray enters the detector with oblique incidence, and the electric vector is parallel to the plane of incidence. To take into account these configurations, we can acquire and analyze data in real working conditions of POLAR-2/LPD.

\subsection{Data collection}

We construct a data collection platform \cite{Xie2023} for calibration of POLAR-2/LPD with linearly polarized X-ray sources at the energy of 4.51 keV. This energy is within the detectable energy range and depends on the anode material (titanium) and the crystal material (LiF). The polarized images from the Topmetal pixel detector are triggered and filtered before passed to storage. Some exemplary images, with three different configurations discussed above, are shown in Fig.~\ref{fig:data}. By carefully examining the images, we can find some notable characteristics:

\begin{itemize}
	\item Almost all images exhibit strong signal features, with a spacially extended track and a bright Bragg peak at one end of the track\footnote{To extract the emission direction of the photoelectron, usually the other end (no Bragg peak) is of interest. Please see the annotations in Fig.~\ref{fig:data}.}.
	\item There is a considerable amount of images, however, leaving tracks in a relatively confined region and showing very subtle directional hints. These images contribute to the randomness in the reconstruction of angular distribution.
	\item Compared to the images with normal incidence, images with oblique incidence display more complicated features, with more dispersed and atypical examples.
\end{itemize}

In general, images show correspondence to major characteristics of the theoretical angular distribution. Extracting the emission angle of the photoelectron can be a challenging task for human experts with direct observations. Besides, it can be difficult to differentiate between configurations with varied incident angles and polarization directions. This makes the data-driven approach appealing in this high-dimensional data analysis task.

\section{Methodology}
\label{sec:method}

\begin{figure*}[htbp]
	\centering
	\includegraphics[width=0.95\textwidth]{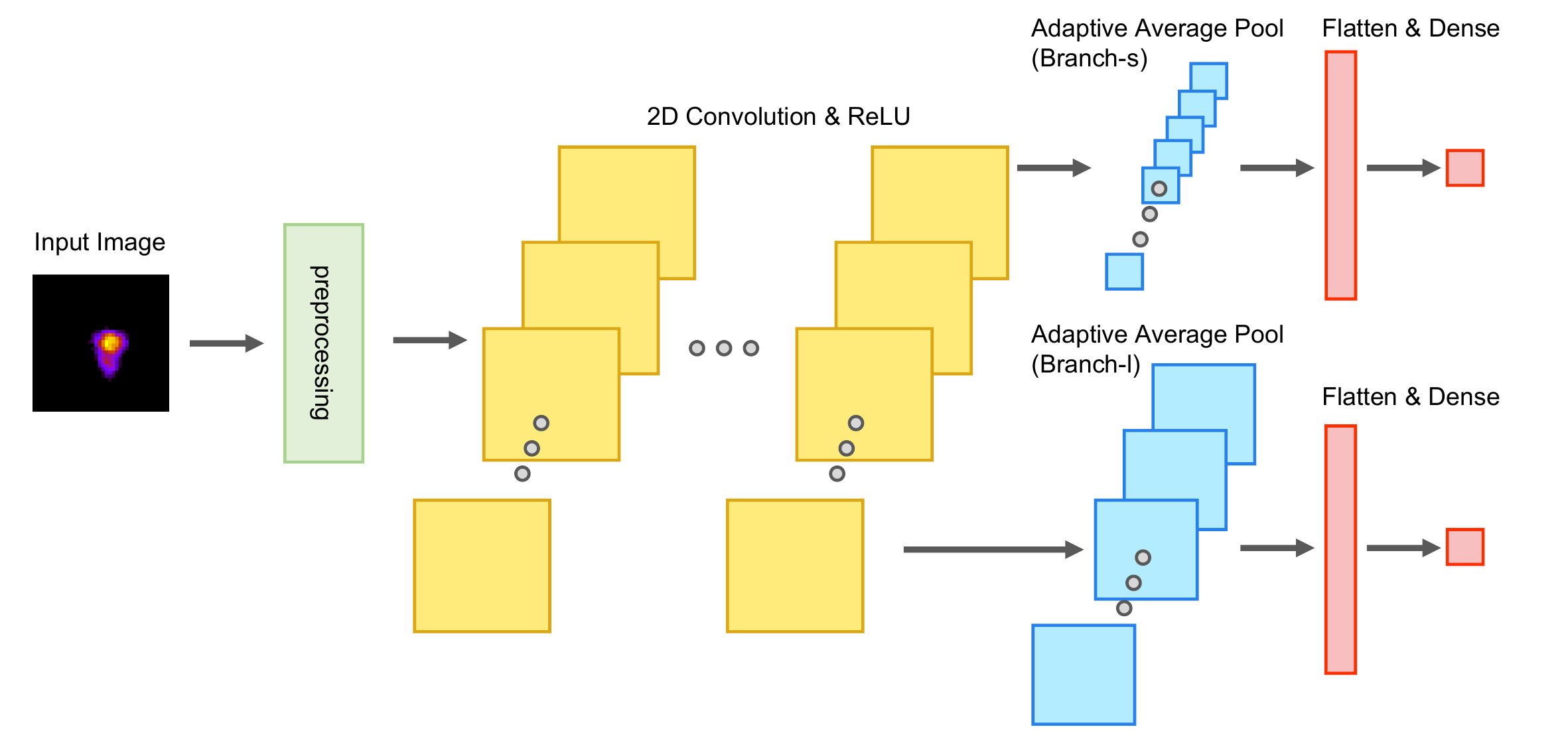}
	\caption{The architecture of individual neural networks for the structured generalized sliced Wasserstein (SGSW) distance. The architecture is composed of 2D convolution, ReLU activation function, adaptive average pooling and dense connections. Choosing different pooling sizes in adaptive average pooling results in different sizes of feature maps before flattening.}\label{fig:nn}
\end{figure*}

\begin{table}[htbp]
	\caption{Details of the neural network architecture of the baseline model.}\label{tab:base-model}%
	\begin{tabular}{@{}llll@{}}
		\toprule
		Part & Name & Kernel Size & Output Size\\
		\midrule
		(Input) & -- & -- & 40$\times$40$\times$1 \\
		\hline
		Backbone & Conv\_1\footnotemark[1] & 3$\times$3$\times$1$\times$32\footnotemark[2] & 40$\times$40$\times$32 \\
		& Conv\_2\footnotemark[1] & 3$\times$3$\times$32$\times$32\footnotemark[2] & 40$\times$40$\times$32 \\
		\hline
		Branch-s\footnotemark[3] & Pool & -- & 1$\times$1$\times$32  \\
		& Flatten & -- & 32 \\
		& Dense & 32$\times$1\footnotemark[2] & 1 \\
		\hline
		Branch-l\footnotemark[3] & Pool & -- & 8$\times$8$\times$32  \\
		& Flatten & -- & 2048 \\
		& Dense & 2048$\times$1\footnotemark[2] & 1 \\
		\botrule
	\end{tabular}
	\footnotetext[1]{The Rectified Linear Units (ReLUs) follow each of these layers.}
	\footnotetext[2]{Here we list the kernel size of weights, the size of biases is equal to the last dimension of the kernel size.}
	\footnotetext[3]{These two branches share the backbone.}
\end{table}

For application of the proposed method, we preprocess the input data for better feature extraction. The original image size is 100$\times$100, we crop it to 40$\times$40 to exclude unfired surrounding pixels. Cropped images are shifted in sub-pixel resolution to locate the center of mass in the middle of the image\footnote{For branch-s of the neural network, center-of-mass aligning can be omitted because this branch is shift-invariant.}. An optional rotation operation is performed before or after the cropping and center-of-mass aligning to study the images with particular polarization directions. Finally, the image is normalized by subtracting the mean intensity and dividing by the standard deviation of the intensity. 

To apply the SGSW distance, we construct the architecture of individual neural networks, illustrated in Fig.~\ref{fig:nn}. The architecture itself is a convolutional neural network commonly found in computer vision tasks \cite{9451544}. What we modify here is the adaptive average pooling to generate outputs with a dual-branch structure: we reduce the feature map to different sizes by this adaptive layer to focus on different scales on the original input data. The two branches assist each other to improve the discrimination power in the generated predictive values after dense connections (see Sect.~\ref{sec:main-res}).

With this architecture, we create a baseline model as our initial attempt for multi-scale feature extraction. The details of the baseline model are given in Table~\ref{tab:base-model}. We use two convolution layers as our backbone, with the Rectified Linear Unit (ReLU) as the activation function. Paddings with zero values are added at the boundary of the feature map to keep its size so that the following adaptive average pooling will be better aligned in pixels. For branch-s, adaptive average pooling reduces the size of the feature map to 1$\times$1 while keeping the channel dimension (32). A densely connected layer further transforms the size to unity after it is flattened. For branch-l, adaptive average pooling reduces the size of the feature map to 8$\times$8. Then, it is flattened to 2048, and transformed to unity by dense connections.

It can be seen that the neural network architecture used here is very lightweight. The model contains only about 11.65k parameters, which is significantly fewer than those of state-of-the-art neural network models. For the GSW distance, models with complicated architectures are not always advantageous for improving discrimination power, because high level features are hard to exploit by deep structures with random weights (see Sect.~\ref{sec:sens} for more discussions).

At initialization, an ensemble of neural networks (64 in this paper) with the same architecture is initialized with random weights and biases. We use the initialization method proposed in \cite{DBLP:conf/iccv/HeZRS15} (as default in the PyTorch \cite{DBLP:conf/nips/PaszkeGMLBCKLGA19} deep learning framework):
\begin{equation} \label{equ:uniform}
	W,\ b \sim \mathcal{U}(-\frac{1}{\sqrt{D}},\ \frac{1}{\sqrt{D}})
\end{equation}

\noindent where:
\begin{equation} \label{equ:d-conv2d}
	D = \mathrm{C}_{\mathrm{in}} \times \mathrm{K}_{\mathrm{height}} \times \mathrm{K}_{\mathrm{width}}
\end{equation}

\noindent for convolution layers, and
\begin{equation} \label{equ:d-linear}
	D = \mathrm{L}_{\mathrm{in}}
\end{equation}

\noindent for linear (densely connected) layers. In equation (\ref{equ:uniform}), $\mathcal{U}(\cdot,\cdot)$ denotes the uniform distribution. In equation (\ref{equ:d-conv2d}), $\mathrm{C}_{\mathrm{in}}$ denotes the input channel size and $\mathrm{K}_{\mathrm{height}/\mathrm{width}}$ denotes the kernel height or kernel width. In equation (\ref{equ:d-linear}), $\mathrm{L}_{\mathrm{in}}$ denotes the input feature length. To ensure neural networks are initialized to various parameters and to ease reproducing the results, we use a series of predefined random seeds for each neural network during initialization.

The Wasserstein-1 distance between two probability distributions ($\mathbb{P}$ and $\mathbb{G}$) is defined by:
\begin{equation}
	W_1 (\mathbb{P}, \mathbb{G}) = \inf\limits_{\gamma \in \prod (\mathbb{P}, \mathbb{G})} (\mathbb{E}_{(x, y) \sim \gamma} ||x - y||^1) \label{equ:wass-def}
\end{equation}

To compute the GSW distance for each branch, we randomly select two batches of examples from two datasets. These two datasets can be polarized images with different configurations, or with different polarization directions achieved by rotation. Suppose we get two batches of data ($\bm{B}_1^{(1:N)}$, $\bm{B}_2^{(1:N)}$) with shape (batch size, image height, image width, image channel), the Wasserstein-1 distance is computed by:
\begin{gather}
	\bm{Y}_{1,i}^{(1:N)} = \mathrm{Model}_i(\bm{B}_1^{(1:N)})\ \ i=1,2,\dots,M \\
	\bm{Y}_{2,i}^{(1:N)} = \mathrm{Model}_i(\bm{B}_2^{(1:N)})\ \ i=1,2,\dots,M \\
	W = \frac{1}{MN} \sum\limits_{i=1}^{M}\sum\limits_{j=1}^N \left| \mathrm{Sort}(\bm{Y}_{1,i})^{(j)} - \mathrm{Sort}(\bm{Y}_{2,i})^{(j)} \right| \label{equ:wass-compute}
\end{gather}

\noindent where $M$ represents the number of neural networks in the ensemble, $N$ represents the number of examples in a batch (batch size), $\mathrm{Model}_i(\cdot)$ stands for the mapping function represented by each neural network, $\mathrm{Sort}(\cdot)$ represents the sorting operation applied to a series of numbers, and equation (\ref{equ:wass-compute}) provides a practical way to compute the Wasserstein-1 distance in equation (\ref{equ:wass-def}) (see Ref. \cite{8578465} for a proof). For SimCLRv2 and DINOv2 in Sect.~\ref{sec:main-res}, equation (\ref{equ:wass-compute}) can still be used if we treat $M$ as the length of the output feature. It can be seen in the following sections that, although each neural network is simple and completely random, the ensemble of neural networks works as a useful and physically informative indicator and gives meaningful distances across datasets, paving the way for effective discrimination.

\section{Results}
\label{sec:res}

We conduct experiments based on the datasets collected by the GPD and the method described above. The datasets include three configurations, in which 168631 examples are for normal incidence (\emph{normal\_inc}), 93856 examples for oblique incidence with the electric vector vertical to the plane of incidence (\emph{oblique\_inc\_v}), and 55834 examples for oblique incidence with the electric vector parallel to the plane of incidence (\emph{oblique\_inc\_p}). Examples are randomly selected from the datasets to form batches. For oblique incidence, the angle between the propagation direction and the normal direction is 30 degrees. The ensemble of neural networks is implemented using the PyTorch \cite{DBLP:conf/nips/PaszkeGMLBCKLGA19} deep learning framework, with 64 individuals in the ensemble. We use graphics processing units (NVIDIA GeForce RTX 4060Ti 16GB) to accelerate the inference of neural networks. We use the baseline model, with batch size of 512, to generate the main results. To improve the statistical significance, we compute the mean and standard deviation based on the results of 100 batches. 

\subsection{Main results}
\label{sec:main-res}

\begin{figure*}[htbp]
	\centering
	\begin{subfigure}{0.49\textwidth}
		\includegraphics[width=\textwidth]{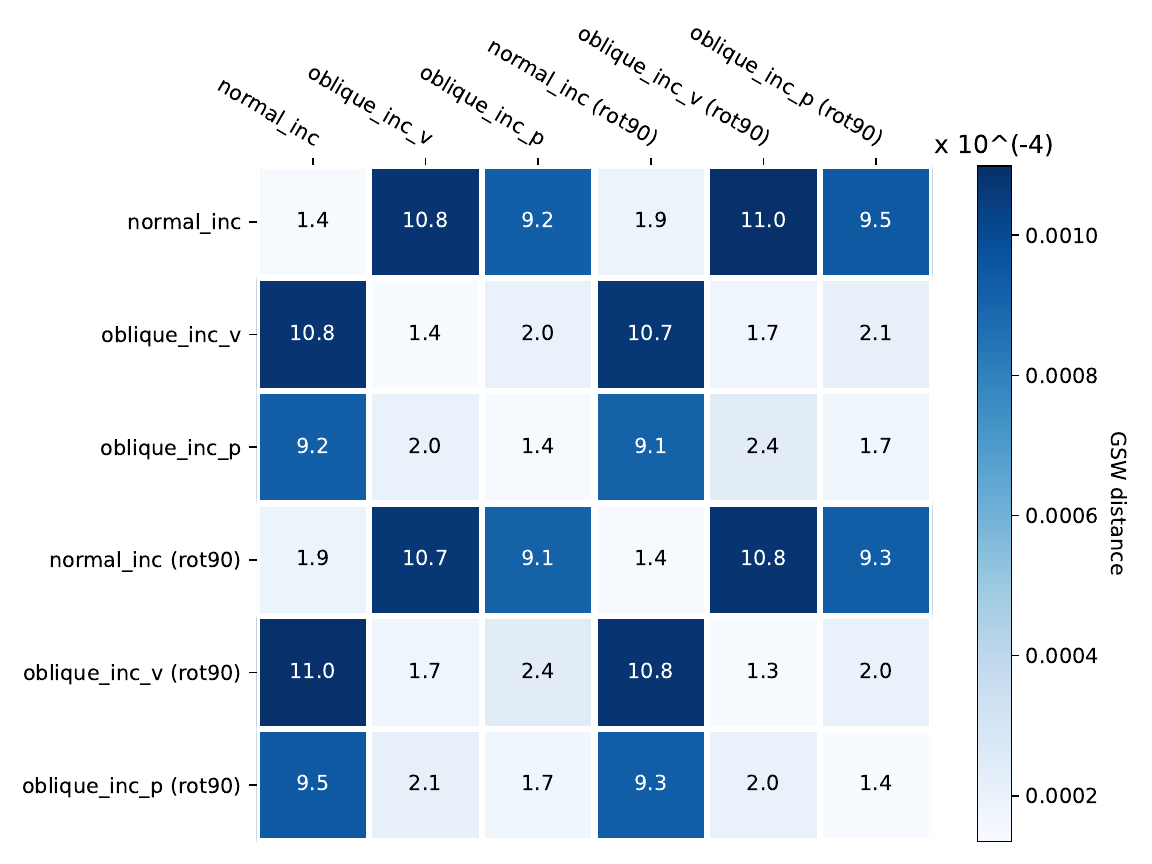}
		\caption{Matrix of branch-s.}
		\label{fig:mat-bs}
	\end{subfigure}
	\hfill
	\begin{subfigure}{0.49\textwidth}
		\includegraphics[width=\textwidth]{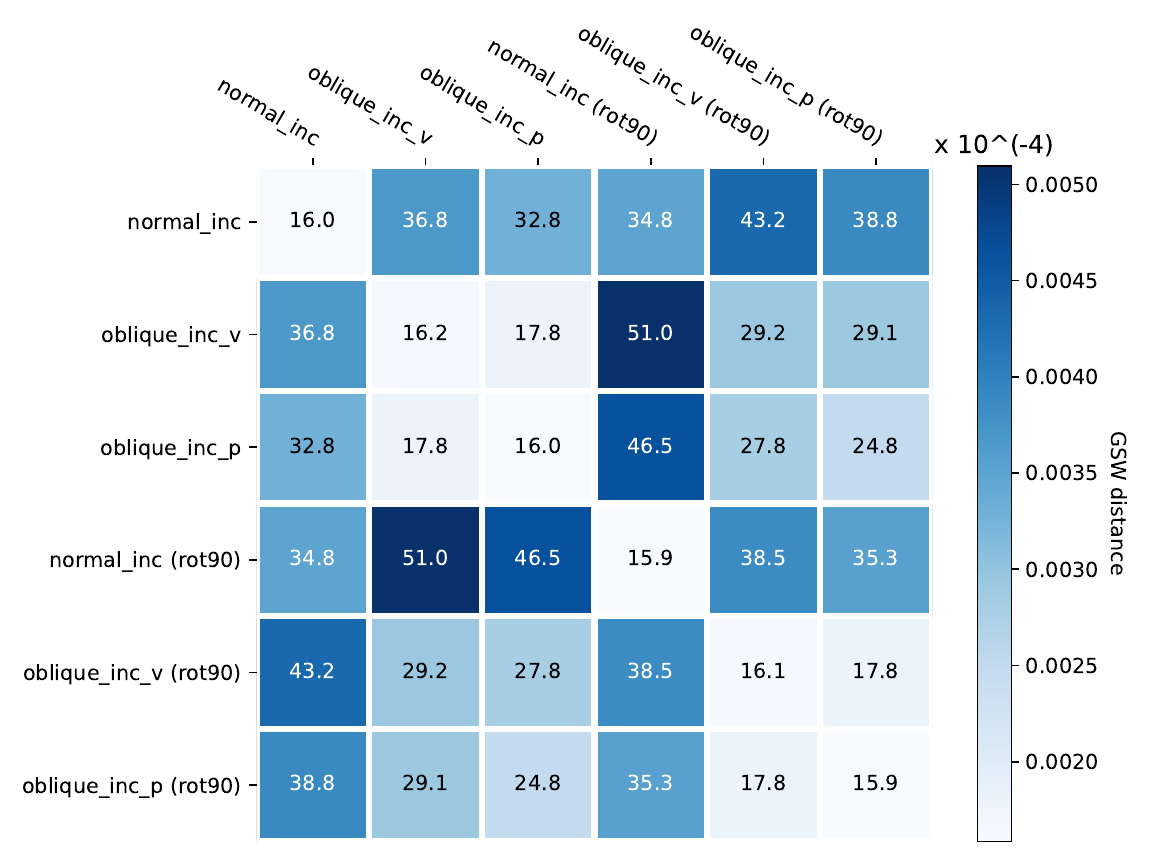}
		\caption{Matrix of branch-l.}
		\label{fig:mat-bl}
	\end{subfigure}
	\caption{Matrices showing the original GSW distances of two branches across different configurations and rotations.}
	\label{fig:mat}
\end{figure*}

\begin{figure*}[htbp]
	\centering
	\begin{subfigure}{0.32\textwidth}
		\includegraphics[width=\textwidth]{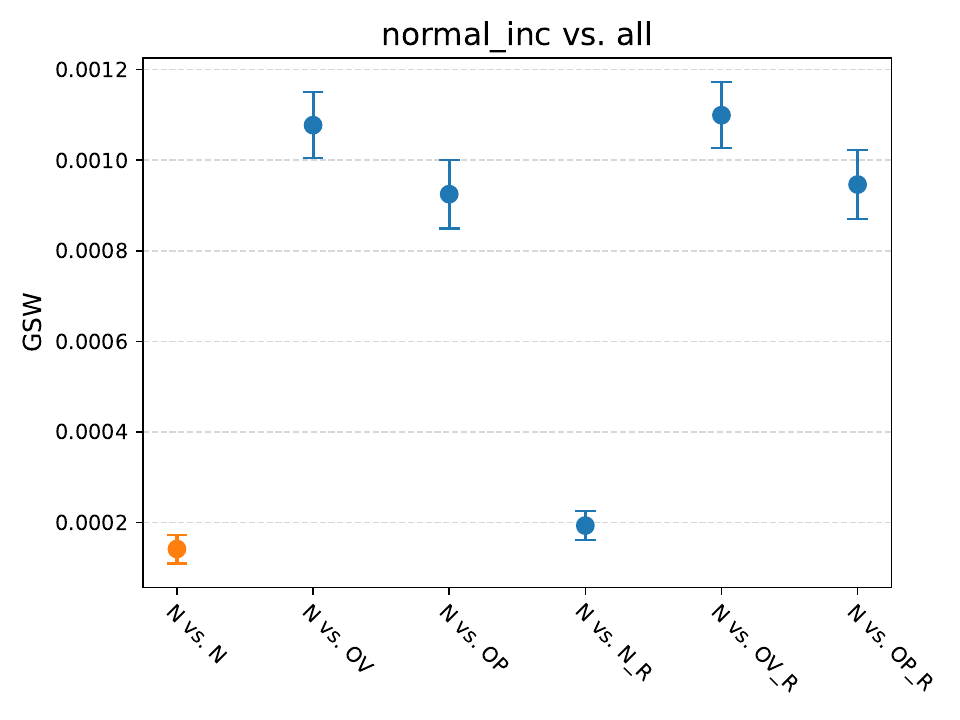}
		\caption{\emph{normal\_inc} in branch-s.}
		\label{fig:norm-bs-unc}
	\end{subfigure}
	\hfill
	\begin{subfigure}{0.32\textwidth}
		\includegraphics[width=\textwidth]{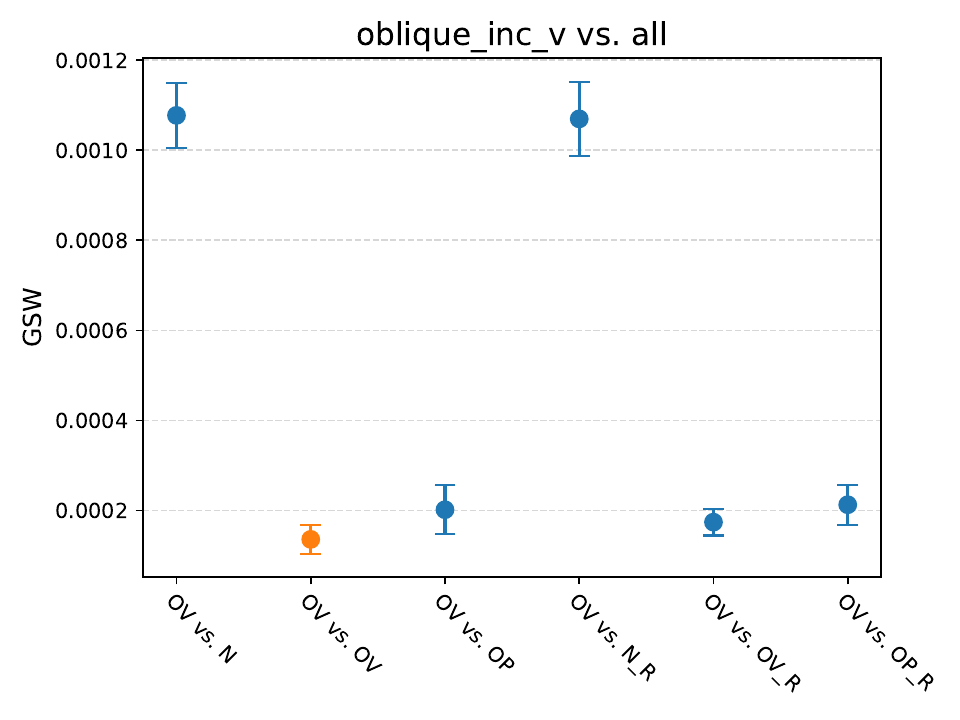}
		\caption{\emph{oblique\_inc\_v} in branch-s.}
		\label{fig:ov-bs-unc}
	\end{subfigure}
	\hfill
	\begin{subfigure}{0.32\textwidth}
		\includegraphics[width=\textwidth]{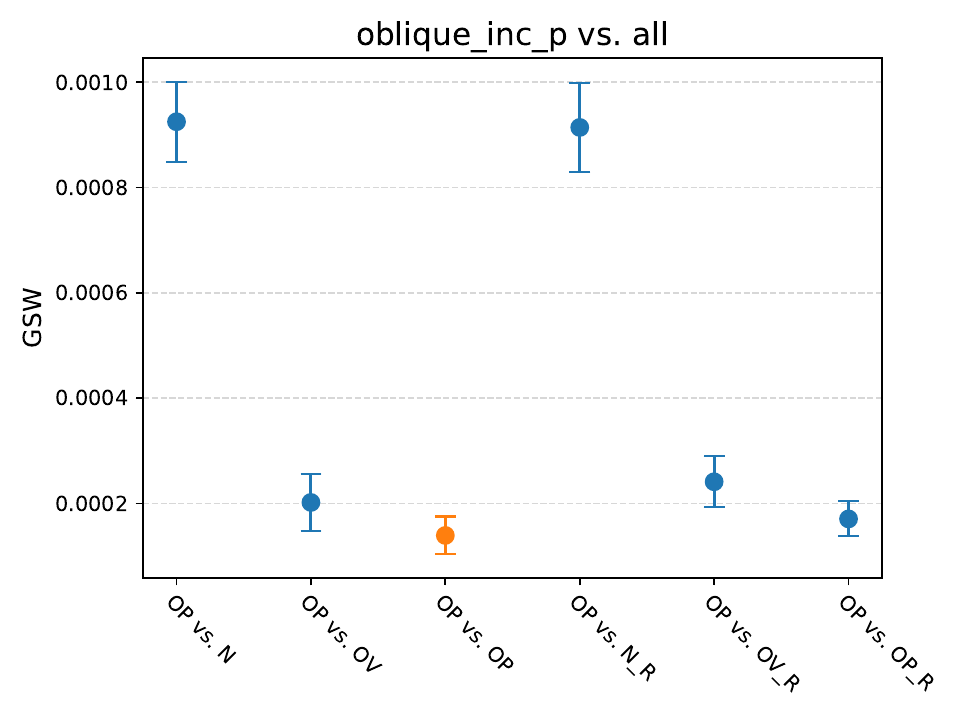}
		\caption{\emph{oblique\_inc\_p} in branch-s.}
		\label{fig:op-bs-unc}
	\end{subfigure}
	\begin{subfigure}{0.32\textwidth}
		\includegraphics[width=\textwidth]{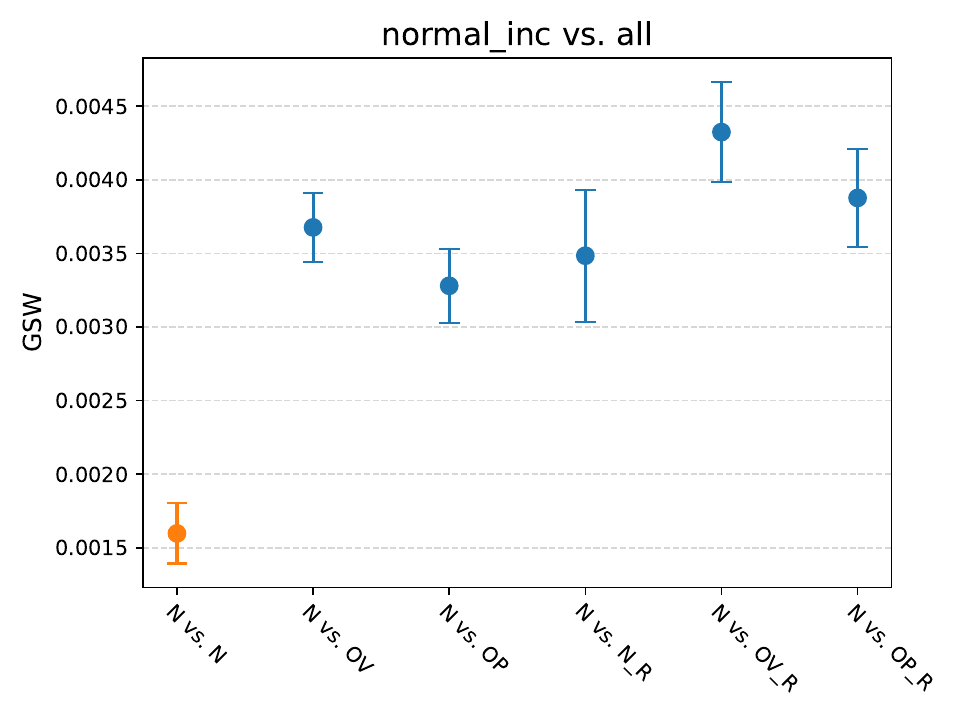}
		\caption{\emph{normal\_inc} in branch-l.}
		\label{fig:norm-bl-unc}
	\end{subfigure}
	\hfill
	\begin{subfigure}{0.32\textwidth}
		\includegraphics[width=\textwidth]{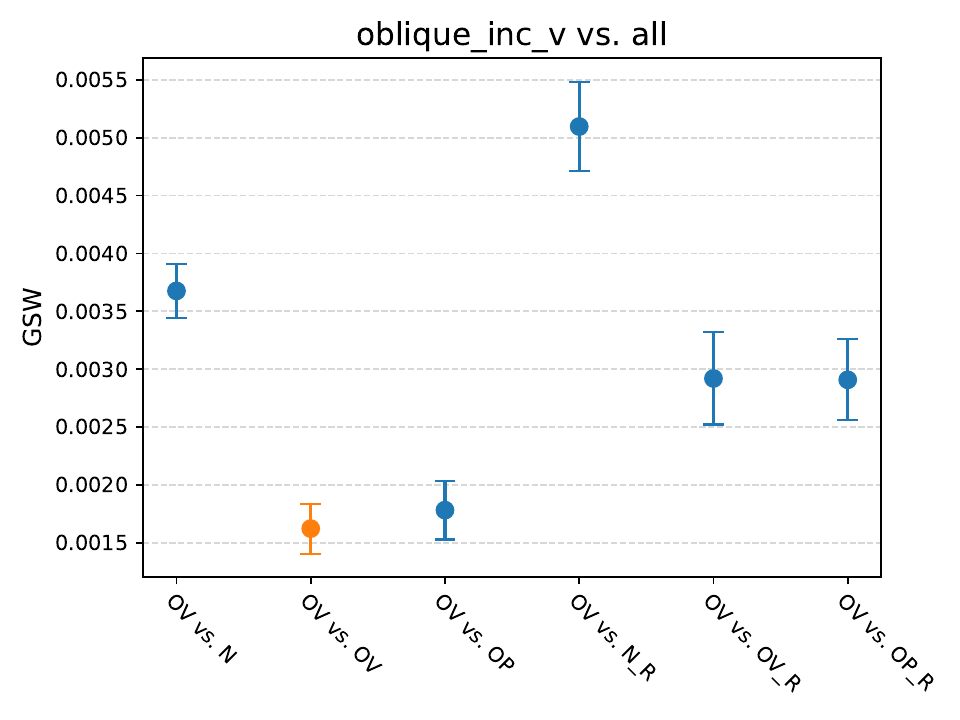}
		\caption{\emph{oblique\_inc\_v} in branch-l.}
		\label{fig:ov-bl-unc}
	\end{subfigure}
	\hfill
	\begin{subfigure}{0.32\textwidth}
		\includegraphics[width=\textwidth]{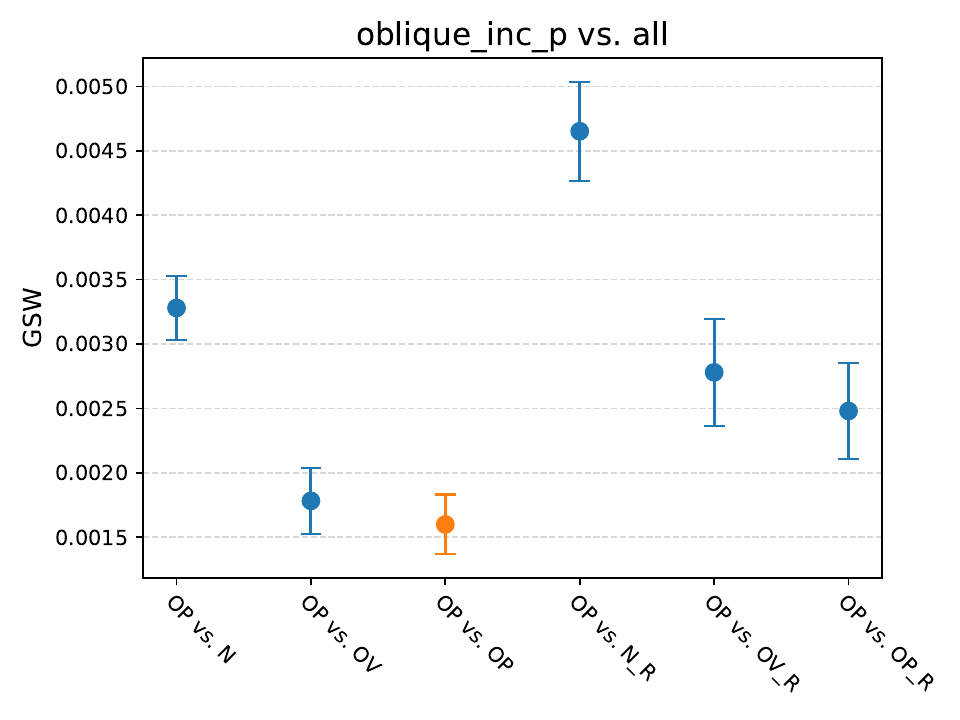}
		\caption{\emph{oblique\_inc\_p} in branch-l.}
		\label{fig:op-bl-unc}
	\end{subfigure}
	\caption{The original GSW distances of two branches with uncertainties (standard deviations) for the top three rows in Fig.~\ref{fig:mat-bs} and Fig.~\ref{fig:mat-bl}. The distance within the same condition is marked with orange, and the distances between different conditions are marked with blue. (N: \emph{normal\_inc}, OV: \emph{oblique\_inc\_v}, OP: \emph{oblique\_inc\_p}, N\_R: \emph{normal\_inc (rot90)}, OV\_R: \emph{oblique\_inc\_v (rot90)}, OP\_R: \emph{oblique\_inc\_p (rot90)}.)}
	\label{fig:unc}
\end{figure*}

Firstly, we evaluate the overall performance of the proposed method using a matrix of GSW distances across several conditions. The matrices showing the original GSW distances are plotted in Fig.~\ref{fig:mat}, where Fig.~\ref{fig:mat-bs} is for the branch-s of the baseline model and Fig.~\ref{fig:mat-bl} is for the branch-l of the baseline model. In total, six conditions are considered: \emph{normal\_inc}, \emph{oblique\_inc\_v}, \emph{oblique\_inc\_p}, and their variations by the 90-degree rotation. The diagonal elements indicate the GSW distances within the same condition, and the non-diagonal elements indicate the GSW distances between different conditions. The numbers in the squares indicate the computed GSW distances.

Compared to the confusion matrix which prefers large values for diagonal elements, the matrix of GSW distances here prefers small values for diagonal elements and large values for non-diagonal elements. It can be seen in Fig.~\ref{fig:mat} that all diagonal elements take the smallest values in their rows and columns, which conforms to the statistical meaning of GSW distances being a metric of data distributions. Besides, the contrast ratio between the non-diagonal element and the corresponding diagonal element indicates the discrimination power of the GSW distances. It can be seen that these two branches have distinct characteristics in discrimination power, indicated by different color patterns.

To more comprehensively evaluate the results, we plot the values of GSW distances with standard deviations for the top three rows in these two matrices. This is shown in Fig.~\ref{fig:unc}. In each sub-figure, we compare one single condition (\emph{normal\_inc}, \emph{oblique\_inc\_v} or \emph{oblique\_inc\_p}) against all conditions. The single condition vs. itself is marked with the color of orange, while others are marked with the color of blue. The strength of branch-s and branch-l can be seen with obvious statistical significance. The branch-s can very well discriminate between different configurations (especially \emph{normal\_inc} vs. \emph{oblique\_inc\_v}/\emph{oblique\_inc\_p}), but does not work as well for rotations; on the other hand, the branch-l can discriminate rotations with noticeable statistical margins. The most difficult case for this model is \emph{oblique\_inc\_v} vs. \emph{oblique\_inc\_p}. For this case, branch-s is better than branch-l, but the statistical margin is not as large as other cases.

\begin{figure*}[htbp]
	\centering
	\begin{subfigure}{0.49\textwidth}
		\includegraphics[width=\textwidth]{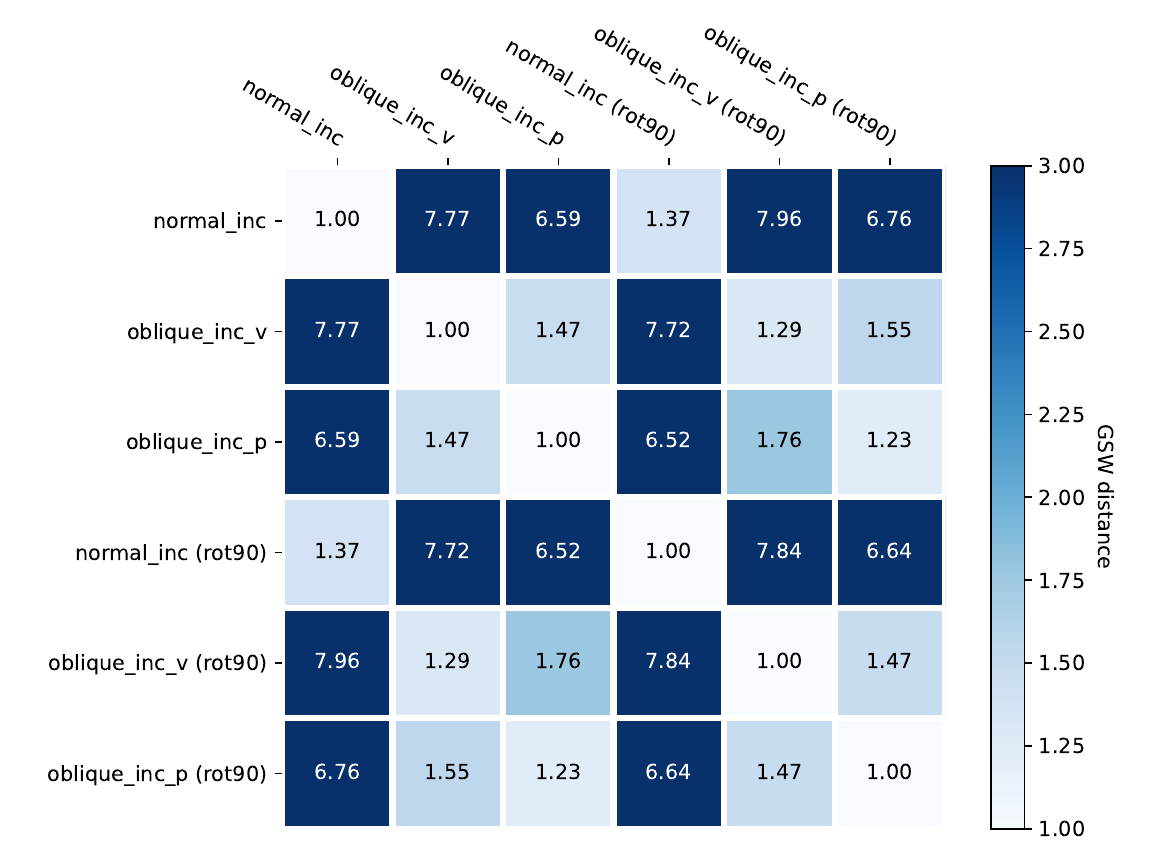}
		\caption{Matrix of branch-s.}
		\label{fig:mat-bs-uni}
	\end{subfigure}
	\hfill
	\begin{subfigure}{0.49\textwidth}
		\includegraphics[width=\textwidth]{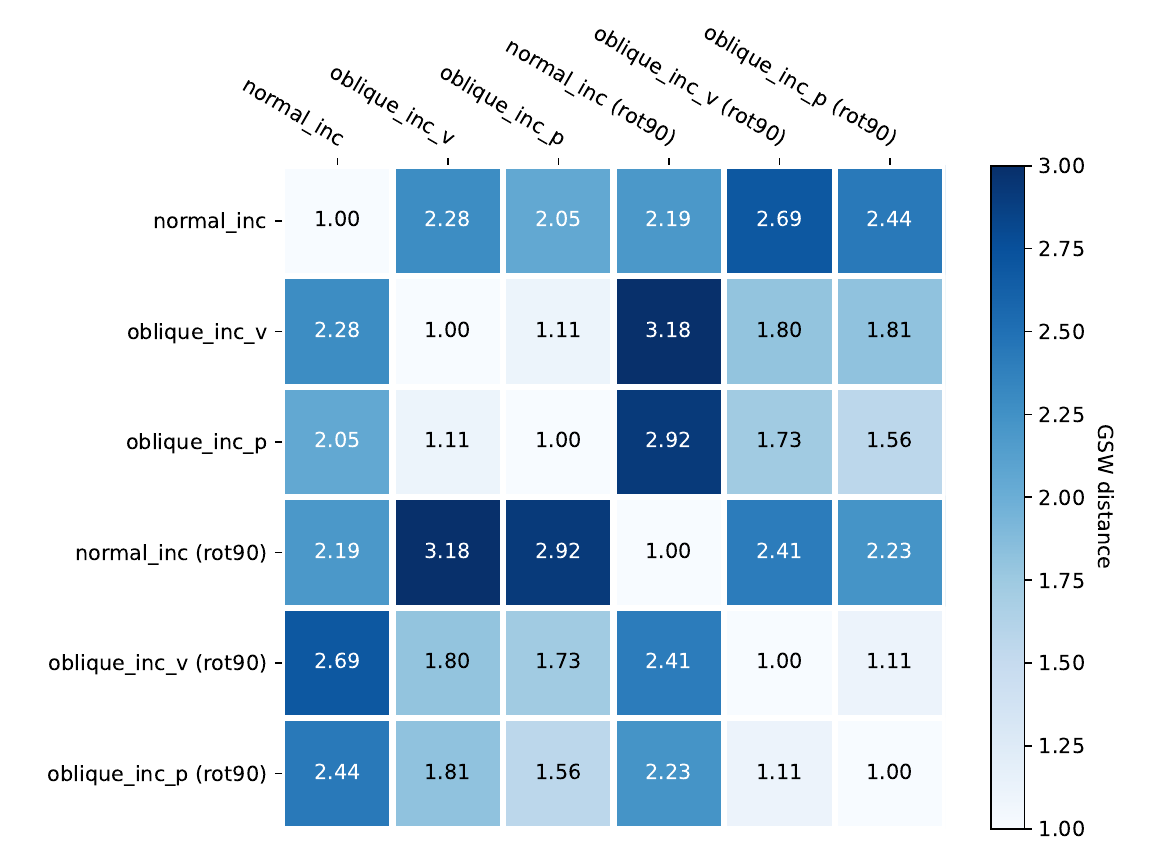}
		\caption{Matrix of branch-l.}
		\label{fig:mat-bl-uni}
	\end{subfigure}
	\caption{Matrices showing the re-normalized GSW distances of two branches across different configurations and rotations. The color bar is for the range from 1.00 to 3.00, and values larger than 3.00 are plotted using the darkest color.}
	\label{fig:mat-uni}
\end{figure*}

\begin{figure*}[htbp]
	\centering
	\includegraphics[width=0.49\textwidth]{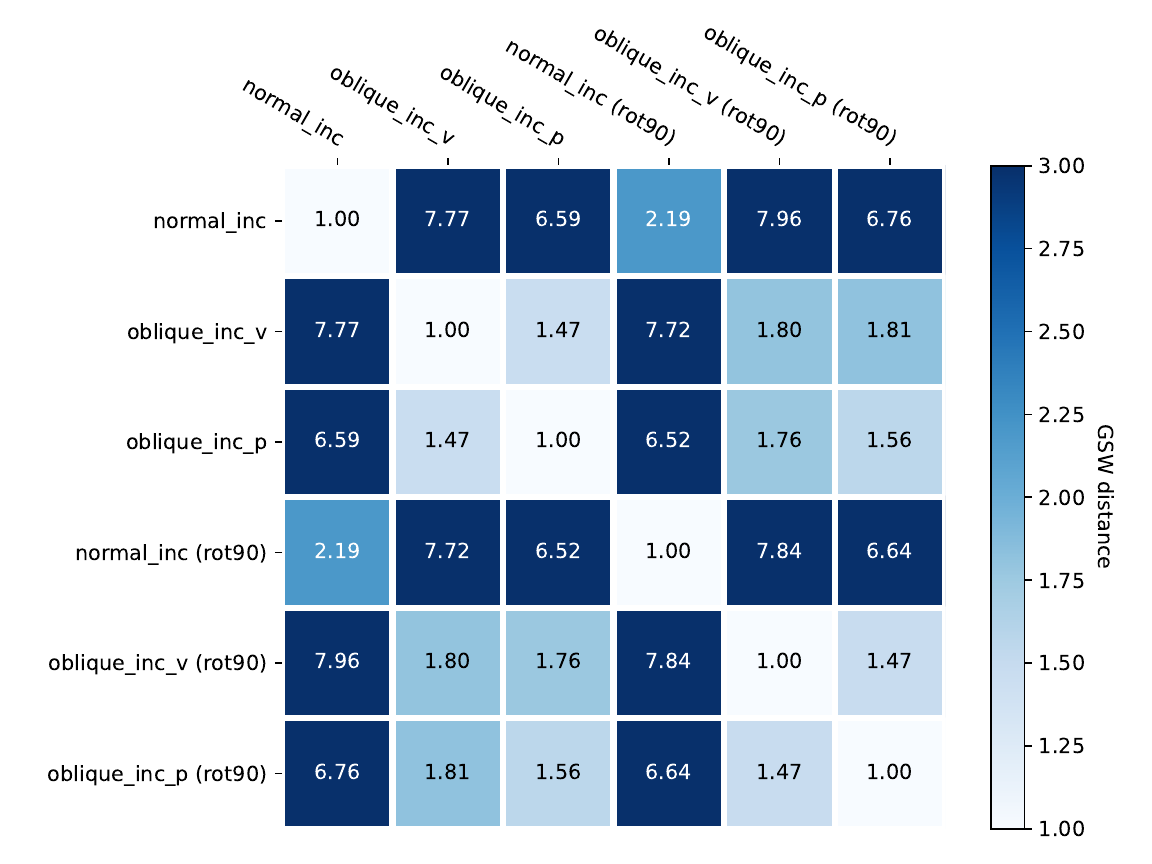}
	\caption{The matrix showing the combined re-normalized GSW distances (SGSW distances) of the two branches.}\label{fig:mat-all-uni}
\end{figure*}

\begin{figure*}[htbp]
	\centering
	\begin{subfigure}{0.49\textwidth}
		\includegraphics[width=\textwidth]{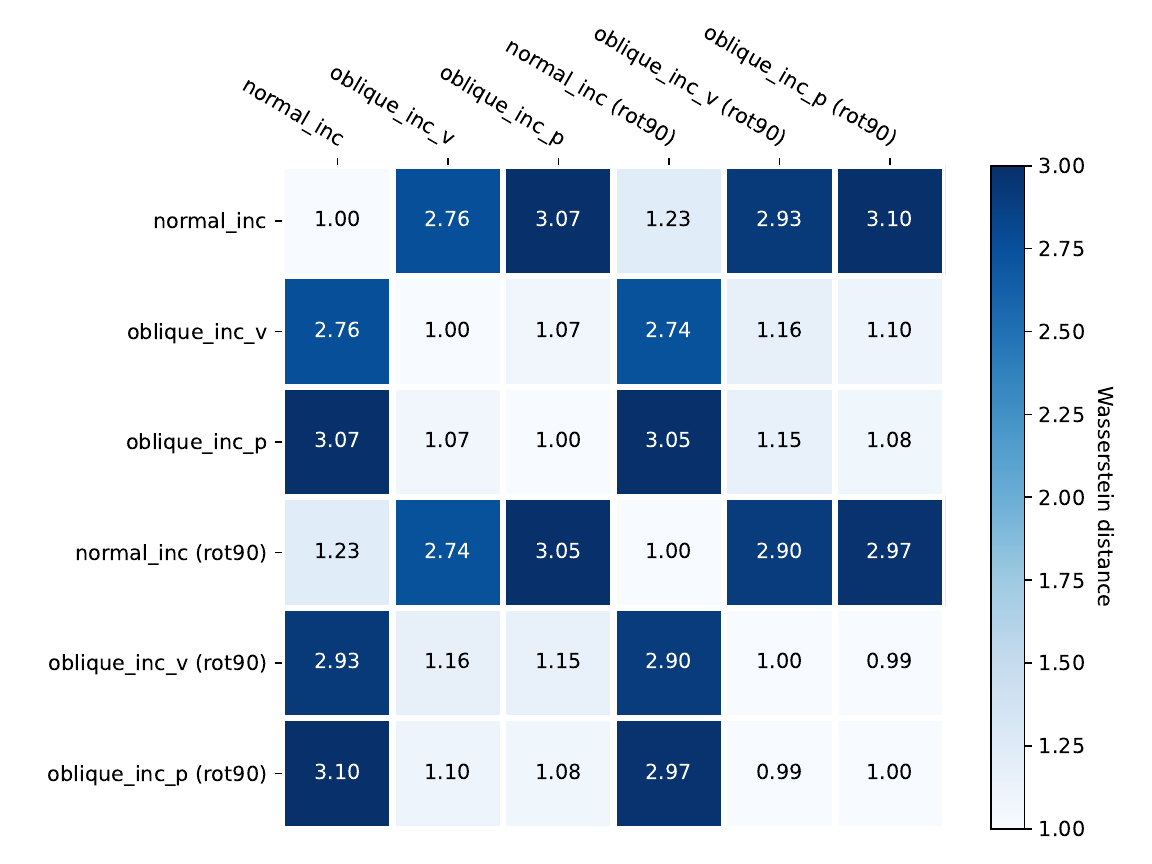}
		\caption{Matrix of SimCLRv2.}
		\label{fig:mat-simclr-uni}
	\end{subfigure}
	\hfill
	\begin{subfigure}{0.49\textwidth}
		\includegraphics[width=\textwidth]{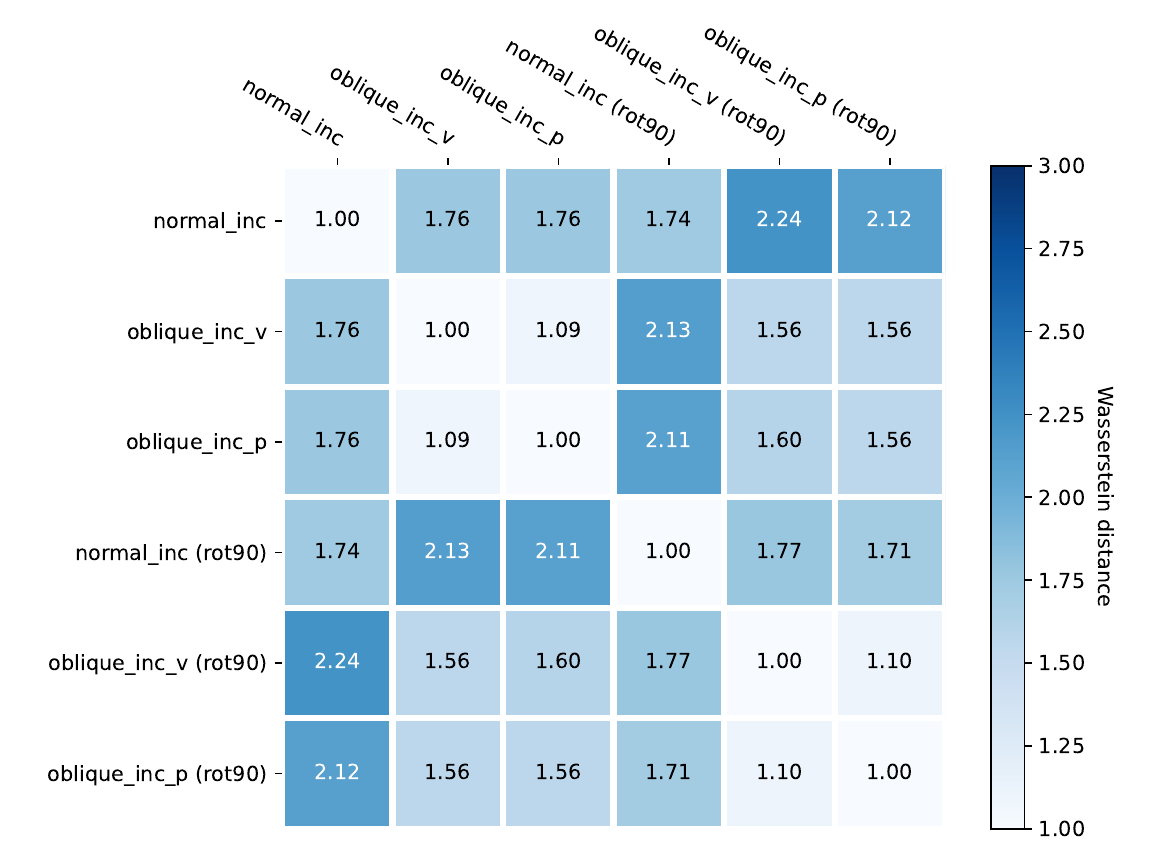}
		\caption{Matrix of DINOv2.}
		\label{fig:mat-dino-uni}
	\end{subfigure}
	\caption{Matrices showing the Wasserstein-1 distances of output features across different configurations and rotations, for self-supervised learning methods. The color bar is for the range from 1.00 to 3.00, and values larger than 3.00 are plotted using the darkest color.}
	\label{fig:mat-self-sup}
\end{figure*}

\begin{figure*}[htbp]
	\centering
	\includegraphics[width=0.9\textwidth]{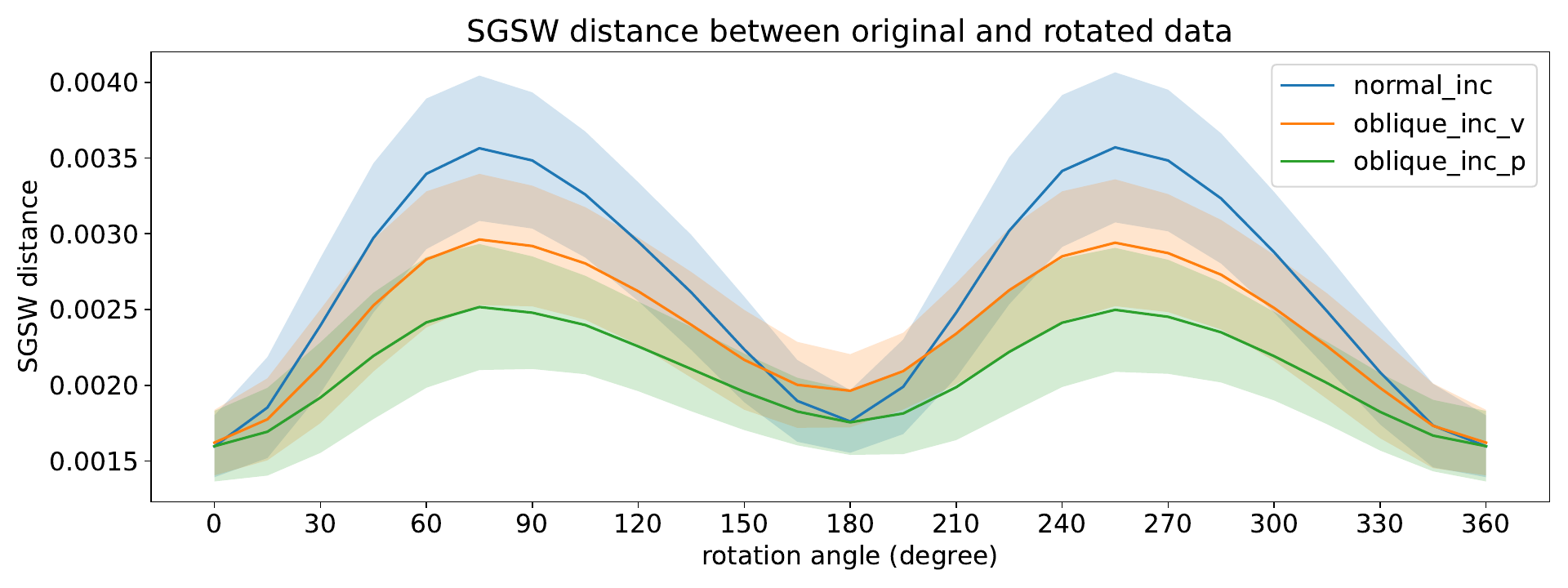}
	\caption{The curves of the SGSW distances between the distribution of original images and the distributions of images after rotation.}\label{fig:rot}
\end{figure*}

\begin{table*}[htbp]
	\caption{Experimental results for all models discussed in Sect.~\ref{sec:main-res} and Sect.~\ref{sec:sens-arch}. The first six rows use the GSW distance for randomized neural networks, and the last two rows use the Wasserstein-1 distance for pre-trained feature extraction networks. (N: \emph{normal\_inc}, OV: \emph{oblique\_inc\_v}, OP: \emph{oblique\_inc\_p}, N\_R: \emph{normal\_inc (rot90)}, OV\_R: \emph{oblique\_inc\_v (rot90)}, OP\_R: \emph{oblique\_inc\_p (rot90)}.)}\label{tab:res-all}
	\setlength{\tabcolsep}{5.5pt}
	\tiny
	\begin{tabular*}{\textwidth}{lccccccccc}
		\hline
		& N vs. N & OV vs. OV & OP vs. OP & N vs. OV & N vs. OP & OV vs. OP & N vs. N\_R & OV vs. OV\_R & OP vs. OP\_R \\
		\hline
		branch-s & 1.00$\pm$0.23 & 1.00$\pm$0.23 & 1.00$\pm$0.26 & \textbf{7.77$\pm$0.52} & \textbf{6.59$\pm$0.54} & \textbf{1.47$\pm$0.39} & 1.37$\pm$0.23 & 1.29$\pm$0.22 & 1.23$\pm$0.24 \\
		branch-l & 1.00$\pm$0.13 & 1.00$\pm$0.13 & 1.00$\pm$0.15 & 2.28$\pm$0.15 & 2.05$\pm$0.16 & 1.11$\pm$0.16 & \textbf{2.19$\pm$0.28} & \textbf{1.80$\pm$0.25} & \textbf{1.56$\pm$0.23} \\
		combined & 1.00$\pm$0.13 & 1.00$\pm$0.13 & 1.00$\pm$0.15 & \textbf{7.77$\pm$0.52} & \textbf{6.59$\pm$0.54} & \textbf{1.47$\pm$0.39} & \textbf{2.19$\pm$0.28} & \textbf{1.80$\pm$0.25} & \textbf{1.56$\pm$0.23} \\
		LeNet & 1.00$\pm$0.12 & 1.00$\pm$0.13 & 1.00$\pm$0.13 & 2.47$\pm$0.17 & 2.17$\pm$0.21 & 1.14$\pm$0.14 & 1.98$\pm$0.26 & 1.69$\pm$0.24 & 1.43$\pm$0.20 \\
		MobileNetV2 & 1.00$\pm$0.04 & 1.00$\pm$0.04 & 1.00$\pm$0.04 & 1.01$\pm$0.04 & 1.01$\pm$0.03 & 1.00$\pm$0.04 & 1.00$\pm$0.04 & 1.01$\pm$0.04 & 1.00$\pm$0.04 \\
		EfficientNet & 1.00$\pm$0.16 & 1.00$\pm$0.22 & 1.00$\pm$0.23 & 1.25$\pm$0.31 & 1.12$\pm$0.23 & 1.00$\pm$0.22 & 1.05$\pm$0.16 & 1.02$\pm$0.21 & 1.02$\pm$0.23 \\
		\hline
		SimCLRv2 & 1.00$\pm$0.26 & 1.00$\pm$0.26 & 1.00$\pm$0.24 & 2.76$\pm$0.51 & 3.07$\pm$0.55 & 1.07$\pm$0.33 & 1.23$\pm$0.21 & 1.16$\pm$0.27 & 1.08$\pm$0.24 \\
		DINOv2 & 1.00$\pm$0.07 & 1.00$\pm$0.07 & 1.00$\pm$0.06 & 1.76$\pm$0.09 & 1.76$\pm$0.10 & 1.09$\pm$0.07 & 1.74$\pm$0.09 & 1.56$\pm$0.09 & \textbf{1.56$\pm$0.08} \\
		\hline
	\end{tabular*}
\end{table*}

The values of GSW distances are on the order of $1 \times 10^{-4}$ for branch-s and $10 \times 10^{-4}$ for branch-l. To evaluate the two branches with a unified scale, we re-normalize the matrices along with the uncertainties by their diagonal elements:
\begin{gather}
	a_{ij} = \frac{1}{K}\sum\limits_{k=1}^{K} W_{ij}^{(k)} \\
	b_{ij} = \sqrt{\frac{\sum_{k=1}^K (W_{ij}^{(k)} - a_{ij})^2}{K}} \\
	\bm{A} = (a_{ij}),\quad \bm{B} = (b_{ij}) \\
	\bm{G} = \mathrm{diag}(\frac{1}{\sqrt{a_{11}}}, \frac{1}{\sqrt{a_{22}}}, \dots, \frac{1}{\sqrt{a_{66}}}) \\
	\bm{\hat{A}} = \bm{G}\bm{A}\bm{G} \\
	\bm{\hat{B}} = \bm{G}\bm{B}\bm{G}
\end{gather}

\noindent where $K$ denotes the number of batches (each batch corresponds to a round of Wasserstein distance computation), $\bm{A}$ and $a_{ij}$ denote the original matrix and its element (mean GSW values), $\bm{B}$ and $b_{ij}$ denote the uncertainty matrix and its element (standard deviations of GSW values), $\bm{G}$ denotes a diagonal matrix, and $\bm{\hat{A}}$, $\bm{\hat{B}}$ denote the re-normalized matrices. After re-normalization, the relative intensities across the matrix are preserved, and all diagonal elements are rendered values of unity. The matrices showing the re-normalized GSW distances are displayed in Fig.~\ref{fig:mat-uni}. This figure more clearly illustrates the discriminative capability of each branch. For branch-s, it can be more definitely discriminated between different configurations (\emph{normal\_inc} vs. \emph{oblique\_inc\_v}, etc.), and less definitely discriminated between rotations (\emph{normal\_inc} vs. \emph{normal\_inc (rot90)}, etc.). On the contrary, for branch-l, it becomes a strength in discrimination between rotations rather than between configurations. Hence, the two branches can work in conditions complementary to each other and jointly improve the discrimination power across configurations and rotations.

To physically interpret the results, we review the neural network architecture of the baseline model. For branch-s, the adaptive average pooling reduces the size of feature map to 1$\times$1. As a result, it focuses on the overall characteristics of the input images but loses most spacially extended information. In our physical context, it can discriminate better between different configurations of incident angles than between rotations of polarized images. For branch-l, the adaptive average pooling reduces the size of feature map to 8$\times$8, so it still keeps the spacially extended information, making it a good discriminator for rotations of polarized images.

In Fig.~\ref{fig:mat-all-uni}, we combine the two re-normalized matrices by taking the larger value in each element:
\begin{equation}
	\tilde{a}_{ij} = \max (\hat{a}_{ij}^s,\ \hat{a}_{ij}^l)
\end{equation}

\noindent where $\tilde{a}_{ij}$ denotes the element of the combined matrix ($\bm{\tilde{A}}$), $\hat{a}_{ij}^s$ denotes the element from the re-normalized matrix of branch-s ($\bm{\hat{A}}^s$), and $\hat{a}_{ij}^l$ denotes the element from the re-normalized matrix of branch-l ($\bm{\hat{A}}^l$). The elements in $\bm{\tilde{A}}$ represent the comprehensive discrimination ability of the proposed SGSW distance. When performing polarization analysis, however, one should choose a particular aspect for discrimination (configuration or rotation) and trace back to a certain branch.

Furthermore, we adapt two commonly used self-supervised learning models to the dataset of polarized images. Unlike SGSW distances which use randomized neural networks, self-supervised learning models, like SimCLRv2 \cite{DBLP:conf/nips/ChenKSNH20} and DINOv2 \cite{DBLP:journals/tmlr/OquabDMVSKFHMEA24}, rely on a pre-training stage to optimize a feature extractor. The feature extractor maps the input image to a feature vector, which is ready for similarity analysis. In Fig.~\ref{fig:mat-self-sup}, we present the re-normalized Wasserstein-1 distances computed upon the output features generated by SimCLRv2 and DINOv2. In Table~\ref{tab:res-all}, we gather the re-normalized results of different models in Sect.~\ref{sec:main-res} and Sect.~\ref{sec:sens-arch} and display their distance metrics with uncertainties. The figure and table indicate, while SimCLRv2 is better at discriminating different configurations, DINOv2 is more sensitive to rotations with less uncertainty than the SGSW method. For example, DINOv2 reaches 1.56$\pm$0.08 for the case of OP vs. OP\_R, while the re-normalized SGSW distance with the combined model is 1.56$\pm$0.23. The reduced uncertainty can be viewed as an advantage when the data samples are limited. For the details of network adaptation, refer to Appendix~\ref{sec:detail-self-sup}.

In analysis of dynamics of polarized lights, it is meaningful to acquire the azimuthal angle of polarization, or the direction of the electric vector, from the angular distribution of photoelectrons. To study the relation between the SGSW distance and the azimuthal angle, we rotate the polarized images in each configuration and compute the SGSW distance between the distributions of original images and the rotated images. The results are gathered as curves in Fig.~\ref{fig:rot}. Three configurations (\emph{normal\_inc}, \emph{oblique\_inc\_v}, \emph{oblique\_inc\_p}) are plotted in separate colors from 0 degree to 360 degrees with a step of 15 degrees, and the error bands indicate the standard deviations.

It can be seen in this figure that all three curves show the correct correspondence to the angular distribution in equation (\ref{equ:angle}). The highest probabilities lie in 0 degree and 180 degrees in the angular distribution, and the lowest probabilities lie in 90 degrees and 270 degrees. These angular characteristics will be reflected by data distributions of polarized images, because the photoelectron ionizes more intensely along its initial emission direction. When rotating through 90 degrees or 270 degrees, the locations of lowest probabilities in the distribution of rotated images coincide with the locations of highest probabilities in the distribution of original images, which gives peaks in the curve of SGSW distances. When rotating through 180 degrees or 360 degrees, the locations of lowest probabilities and highest probabilities do not change, which gives valleys in the curve of SGSW distances. Although not perfectly aligning with the theoretically analysis, the curves have similar trends and distinct peaks and valleys. Besides, the configuration \emph{normal\_inc} has the highest peak-to-valley ratio, with the configuration \emph{oblique\_inc\_v} being the second and the configuration \emph{oblique\_inc\_p} being the third. An explanation of this phenomenon by a simplified statistical model is given in Sect.~\ref{sec:stat}.

\subsection{Analysis of sensitivity}
\label{sec:sens}

\subsubsection{Sensitivity to architectures}
\label{sec:sens-arch}

In this section, we investigate alternative network architectures and judge their performance under the same GSW framework. We experiment with variants of the LeNet \cite{726791}, MobileNetV2 \cite{8578572} and EfficientNet \cite{DBLP:conf/icml/TanL19}. The network architectures are adapted to our image size (40$\times$40) and single channel. We use the same initialization method for network weights as discussed in Sect.~\ref{sec:method}. The results are re-normalized and compared to other models in Table~\ref{tab:res-all}. It can be seen that the LeNet variant has demonstrate good discrimination power especially for cases with rotations, but MobileNetV2 and EfficientNet variants do not show sufficient discrimination abilities. See Sect.~\ref{sec:detail-sens-arch} for details of these network architectures.

To explain why classic architectures in computer vision do not work as desired under the GSW framework, we give two major reasons: the data pattern and the nature of randomized neural networks. First, although the image size is 40$\times$40, most active pixels locate in a relatively confined center region after preprocessing, so generally we do not need many stacked layers (Convolution and Pool) like LeNet to have a large receptive field. This is in contrast to the MNIST dataset and many other computer vision datasets, which have global features and abundant translation/rotation variations. Second, since the weights of the neural networks are randomly initialized, their performance does not rely on the optimization process as usually in the supervised/self-supervised setting. Some elaborated architectures, like MobileNetV2 or EfficientNet, might rely on the gradient-based optimization process for proper functioning, such as pre-training as a backbone for other tasks. As a result, the optimal architecture for computing the GSW distance is determined by how efficiently we can acquire information from image data with randomized neural networks, and sometimes is subject to particular considerations.

\subsubsection{Sensitivity to hyperparameters}

\begin{figure*}[htbp]
	\centering
	\begin{subfigure}{0.45\textwidth}
		\includegraphics[width=\textwidth]{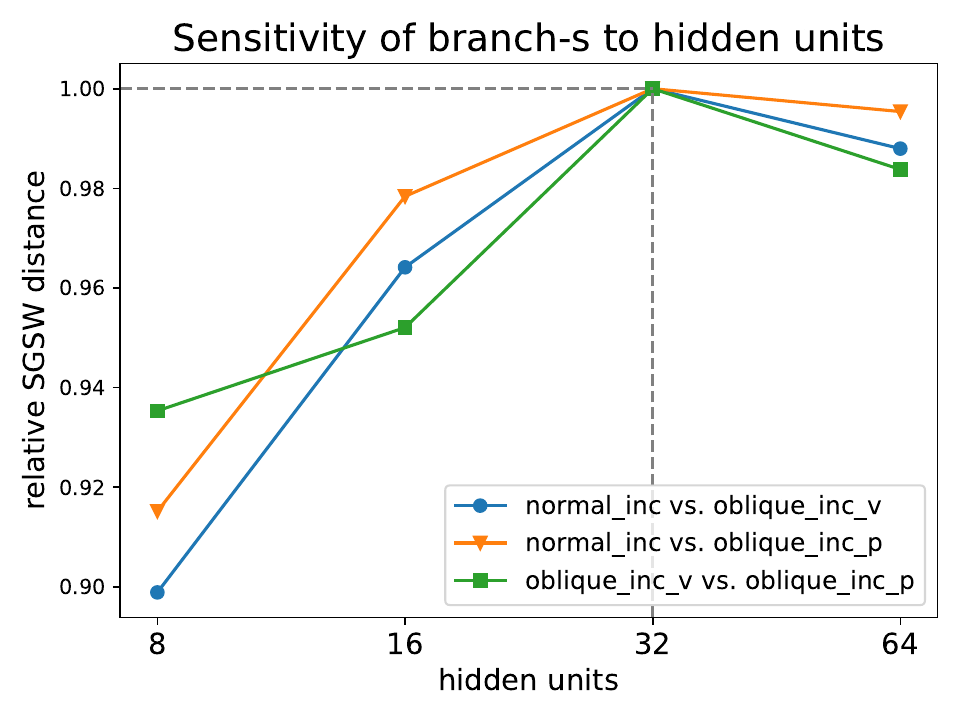}
		\caption{Hidden units for branch-s.}
		\label{fig:v1-hidden}
	\end{subfigure}
	\hfill
	\begin{subfigure}{0.45\textwidth}
		\includegraphics[width=\textwidth]{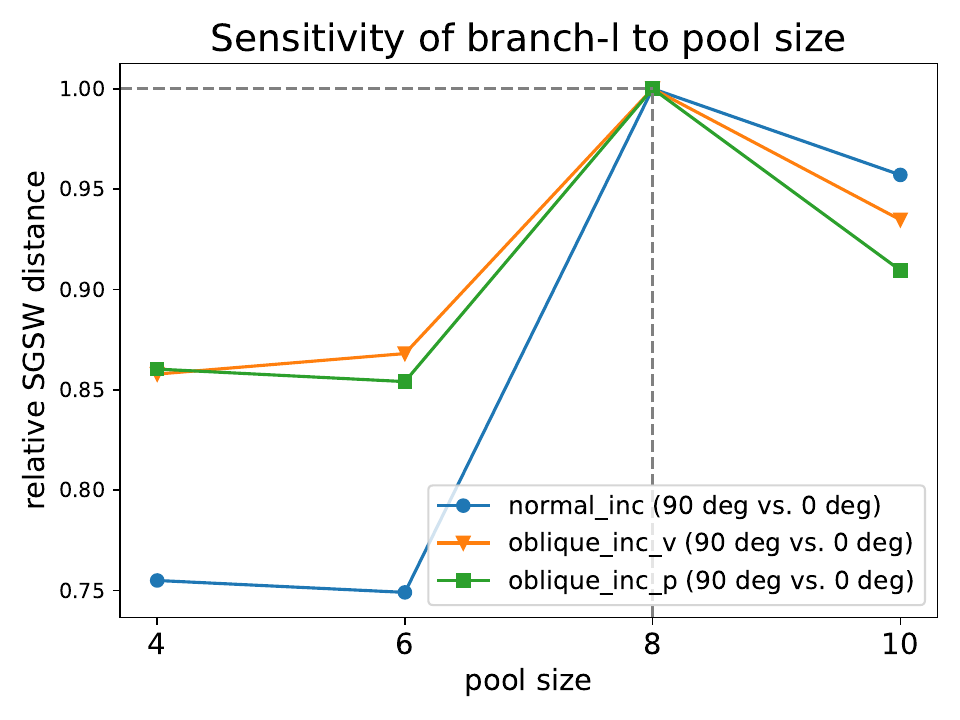}
		\caption{Pool size for branch-l.}
		\label{fig:v2-pool}
	\end{subfigure}
	\hfill
	\begin{subfigure}{0.45\textwidth}
		\includegraphics[width=\textwidth]{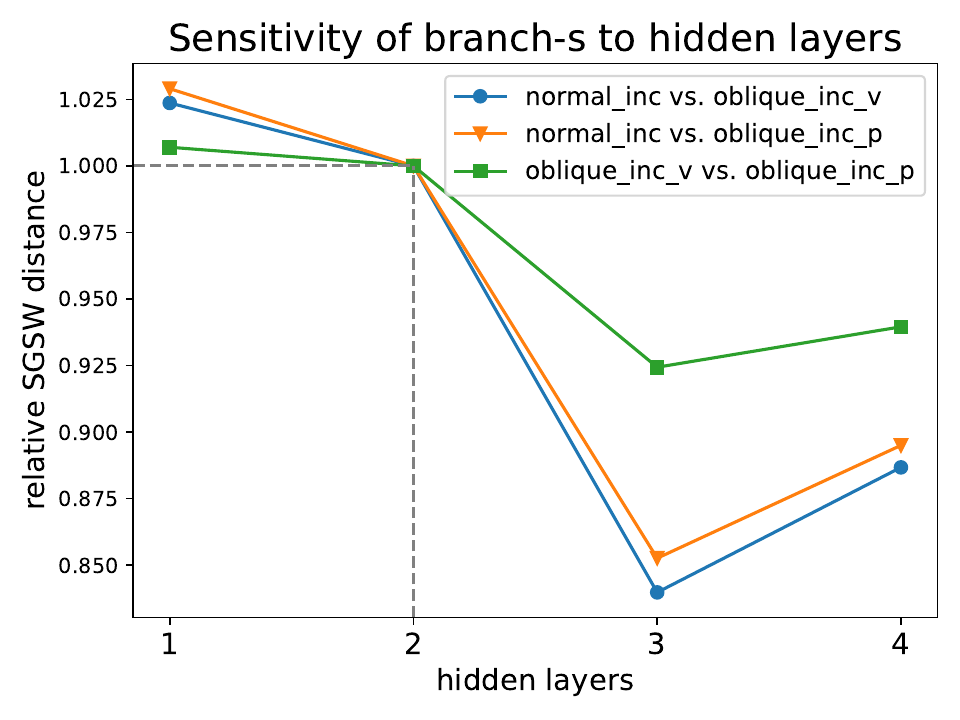}
		\caption{Layers for branch-s.}
		\label{fig:v1-layer}
	\end{subfigure}
	\hfill
	\begin{subfigure}{0.45\textwidth}
		\includegraphics[width=\textwidth]{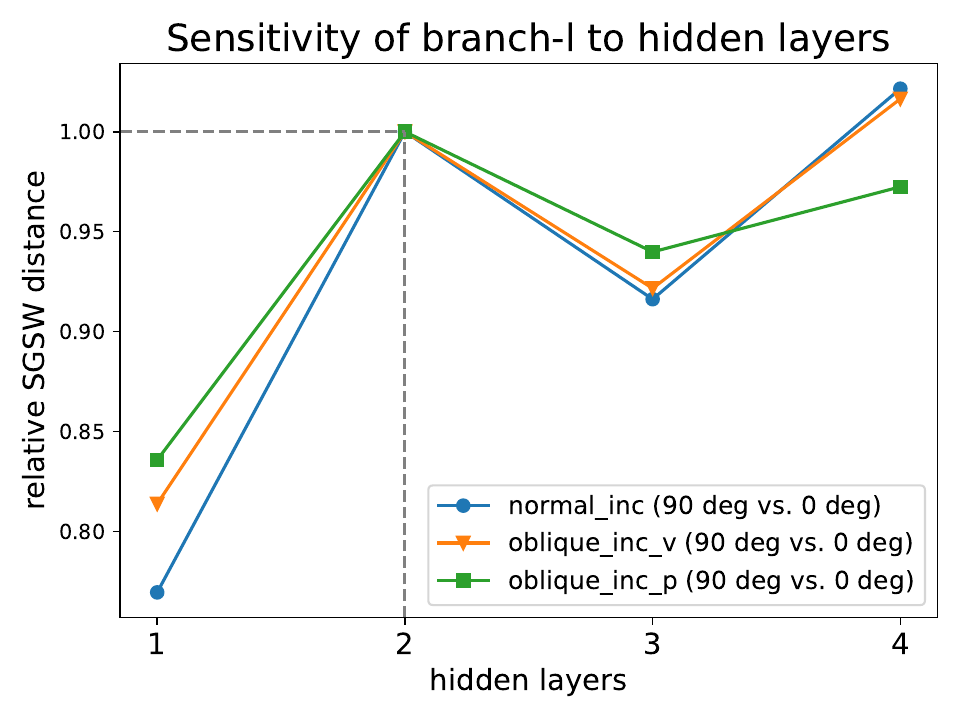}
		\caption{Layers for branch-l.}
		\label{fig:v2-layer}
	\end{subfigure}
	\hfill
	\begin{subfigure}{0.45\textwidth}
		\includegraphics[width=\textwidth]{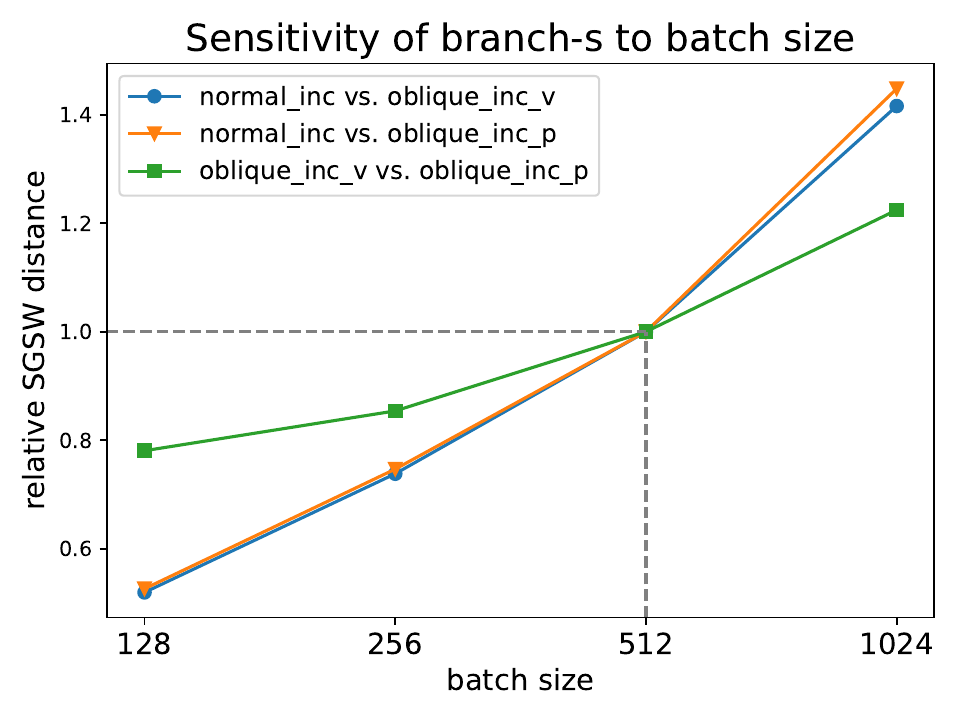}
		\caption{Batch size for branch-s.}
		\label{fig:v1-batch}
	\end{subfigure}
	\hfill
	\begin{subfigure}{0.45\textwidth}
		\includegraphics[width=\textwidth]{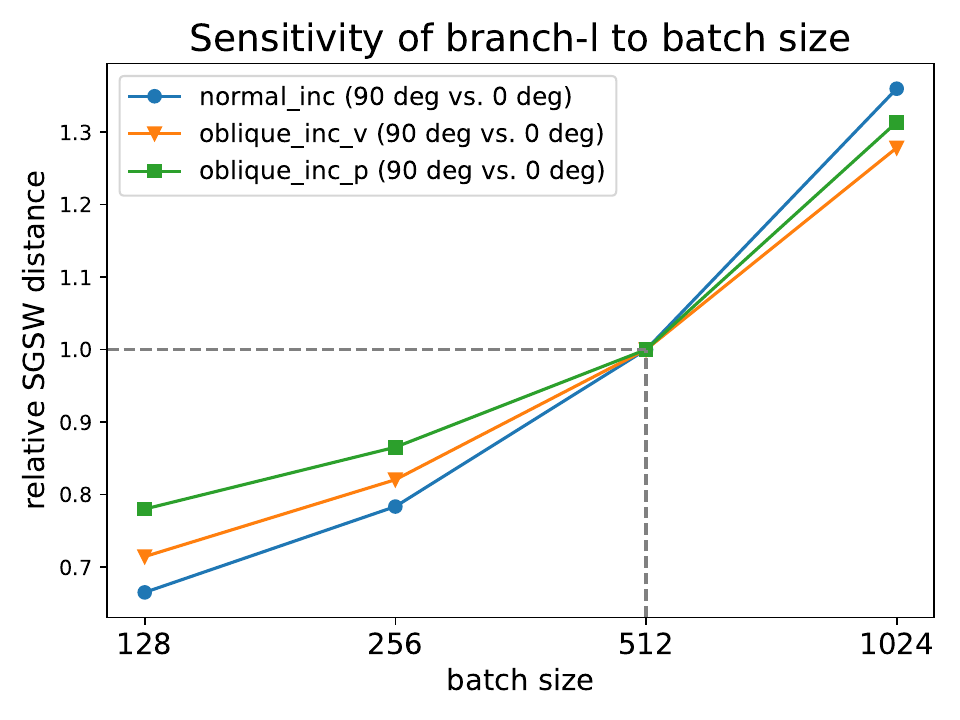}
		\caption{Batch size for branch-l.}
		\label{fig:v2-batch}
	\end{subfigure}
	\caption{Line plots showing the sensitivity of two branches to hyperparameters of the neural network and the batch size. We use the baseline model as the reference when comparing different values of variables.}
	\label{fig:sens}
\end{figure*}

\begin{figure*}[htbp]
	\centering
	\includegraphics[width=0.95\textwidth]{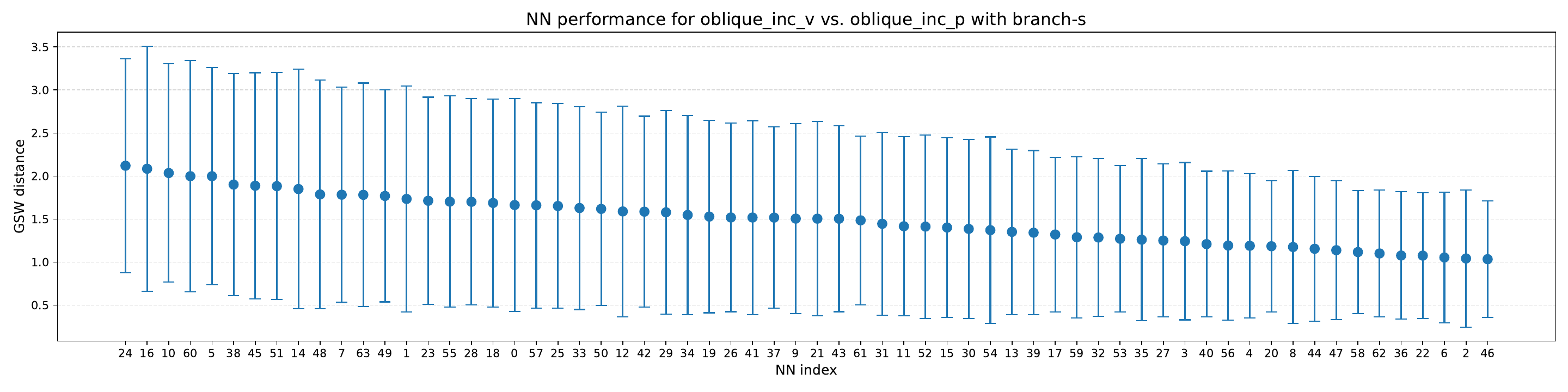}
	\caption{Re-normalized GSW distances of individual networks in the ensemble. This figure shows the performance of branch-s for the case \emph{oblique\_inc\_v} vs. \emph{oblique\_inc\_p}.}\label{fig:perf-v-vs-p}
\end{figure*}

\begin{figure*}[htbp]
	\centering
	\includegraphics[width=0.95\textwidth]{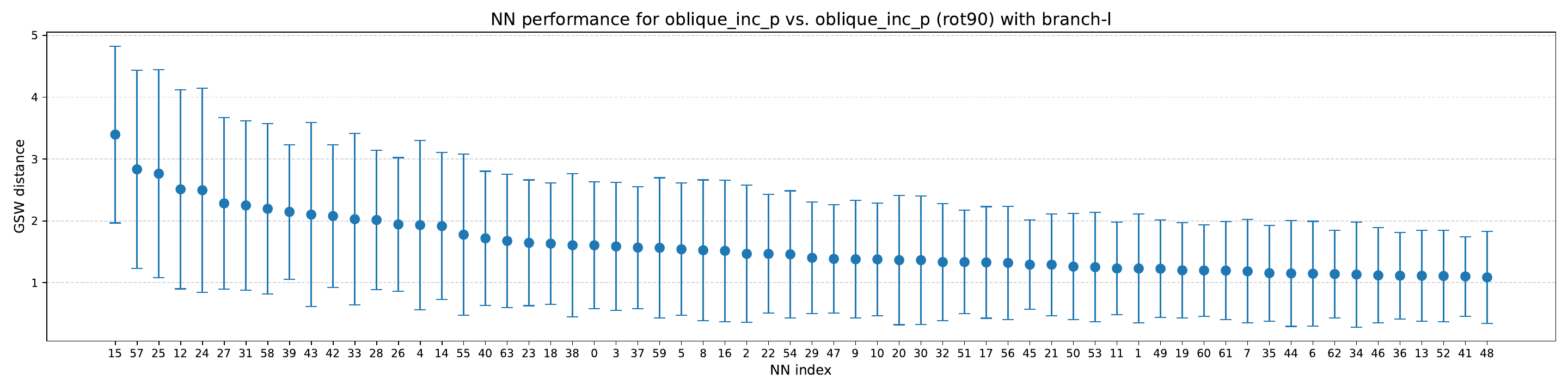}
	\caption{Re-normalized GSW distances of individual networks in the ensemble. This figure shows the performance of branch-l for the case \emph{oblique\_inc\_p} vs. \emph{oblique\_inc\_p (rot90)}.}\label{fig:perf-p-vs-p90}
\end{figure*}

\begin{figure*}[htbp]
	\centering
	\begin{subfigure}{0.49\textwidth}
		\includegraphics[width=\textwidth]{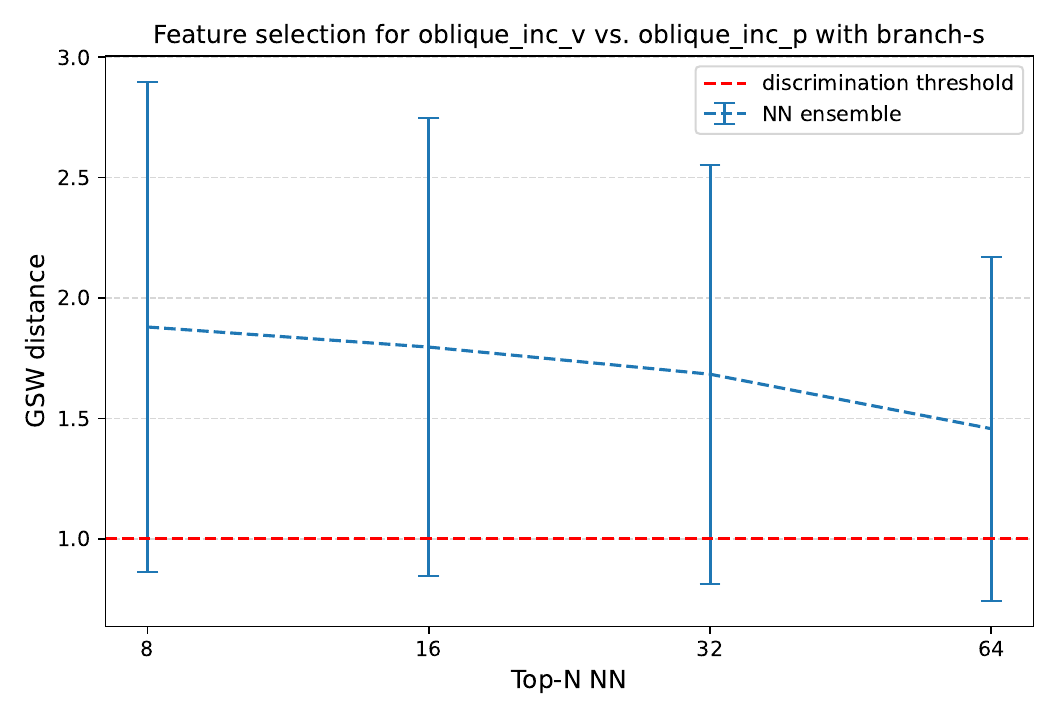}
		\caption{Feature selection for branch-s.}
		\label{fig:feat-sel-s}
	\end{subfigure}
	\hfill
	\begin{subfigure}{0.49\textwidth}
		\includegraphics[width=\textwidth]{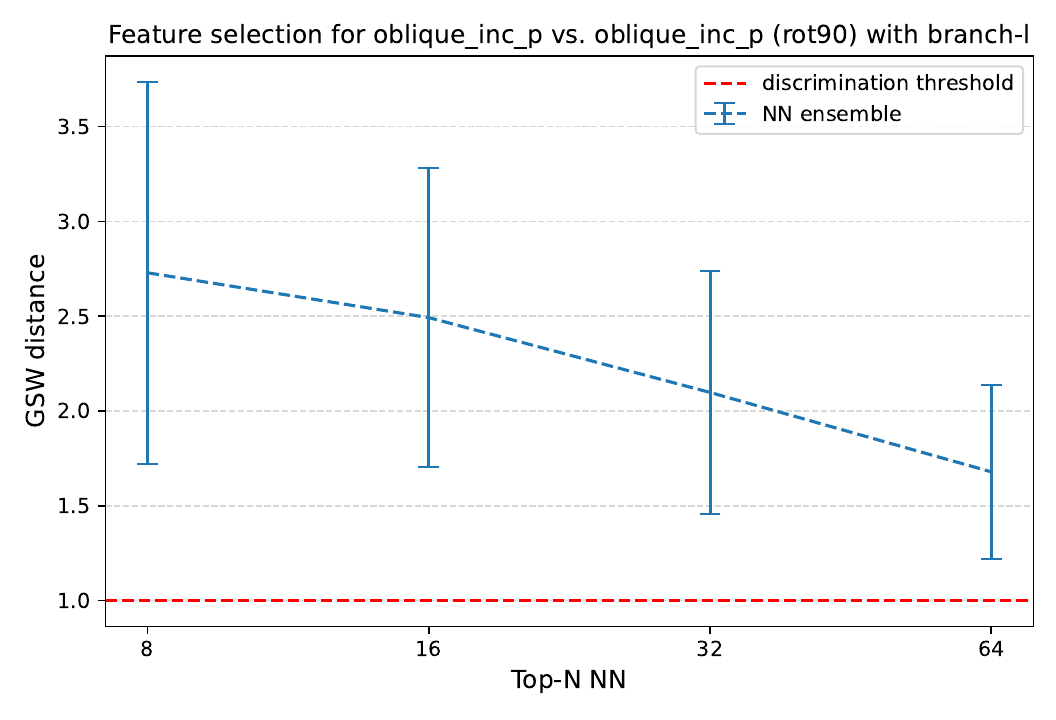}
		\caption{Feature selection for branch-l.}
		\label{fig:feat-sel-l}
	\end{subfigure}
	\caption{The re-normalized GSW distances with different numbers of individual networks in the ensemble. We select the top-n networks with the highest mean values of re-normalized GSW distances.}
	\label{fig:feat-sel}
\end{figure*}

To investigate the influence of changing neural network hyperparameters within the architecture and number of examples in the distribution, we perform a sensitivity analysis on hyperparameters and the batch size. The results are gathered in Fig.~\ref{fig:sens}. For branch-s, we conduct experiments between different configurations. For branch-l, we conduct experiments between each configuration and its rotated variation. All results are re-normalized and relative to the baseline model. The discussions of the results are presented as follows:

\begin{itemize}
	\item In Fig.~\ref{fig:v1-hidden}, we investigate the impact of hidden units on the discrimination performance of branch-s. We find that, increasing the number of hidden units initially improves the the discrimination performance (from 8 to 32), but when hidden units reach a certain value, the performance begins to shrink. It indicates that there is an optimal choice of hidden units to restrict the complexity of the neural network.
	\item In Fig.~\ref{fig:v2-pool}, we investigate how the size of adaptive average pooling impacts the discrimination performance of branch-l. When the pool size is small (below 6), the performance is limited. When the pool size increases to a certain amount, the performance is boosted, but begins to shrink when increased further. It indicates the reasonable choice of the pool size is important for discrimination performance of branch-l.
	\item In Fig.~\ref{fig:v1-layer} and Fig.~\ref{fig:v2-layer}, we investigate the impact of hidden convolution layers on the discrimination power. For branch-s, fewer layers are more preferable to improve the performance. However, for branch-l, a large gap exists between one hidden layer and two or more hidden layers. It indicates, while branch-s can better discriminate on shallow features, branch-l needs the deep structure to reach the ideal performance.
	\item In Fig.~\ref{fig:v1-batch} and Fig.~\ref{fig:v2-batch}, we investigate the impact of the batch size on the discrimination power. In can be seen that, the performance of two branches increases monotonically with the batch size. This result is reasonable considering the data distribution of polarized images with a substantial amount showing subtle directional hints. More examples will illustrate the distributional profile more accurately, and thus give larger SGSW distances.
\end{itemize}

\subsection{Feature selection}

From the perspective of feature engineering, each individual network in the ensemble provides a feature by random projections. Thus, it is possible to select prominent features to reduce computation and to improve the model performance. Two most difficult cases (\emph{oblique\_inc\_v} vs. \emph{oblique\_inc\_p} and \emph{oblique\_inc\_p} vs. \emph{oblique\_inc\_p (rot90)}) are chosen as pilot experiments. We use the first half of the dataset in each configuration to select prominent features, and the second half to test the performance of selected features.

In Fig.~\ref{fig:perf-v-vs-p} and Fig.~\ref{fig:perf-p-vs-p90}, we sort individual networks by the mean values of their re-normalized GSW distances in descending order. The uncertainties (standard deviations) of the numerator and the denominator are merged during re-normalization to account for both with one index. The diversity of performance of individual networks can be clearly recognized. The individuals with larger mean values tend to have larger standard deviations, with more discrimination power as a whole. For the case \emph{oblique\_inc\_v} vs. \emph{oblique\_inc\_p} with branch-s, the re-normalized GSW distances range from 2.12$\pm$1.24 to 1.03$\pm$0.68; for the case \emph{oblique\_inc\_p} vs. \emph{oblique\_inc\_p (rot90)} with branch-l, the re-normalized GSW distances range from 3.40$\pm$1.43 to 1.09$\pm$0.75.

In Fig.~\ref{fig:feat-sel-s} and Fig.~\ref{fig:feat-sel-l}, we show the discrimination capabilities by selecting top-n individual networks in the ensemble. We select top-8, top-16, top-32 and all individuals and plot the sub-figures. Although suffering from the risk of increasing uncertainties, the feature selection method does improve the discrimination performance to some extent. The re-normalized GSW distances have improved from 1.46$\pm$0.71 to 1.88$\pm$1.02, and from 1.68$\pm$0.46 to 2.73$\pm$1.01, in these two cases. However, the improvement should be objectively regarded, since using fewer individuals also shrinks the robustness of the ensemble model.

\subsection{Predictive performance}

\begin{figure}[htbp]
	\centering
	\includegraphics[width=0.48\textwidth]{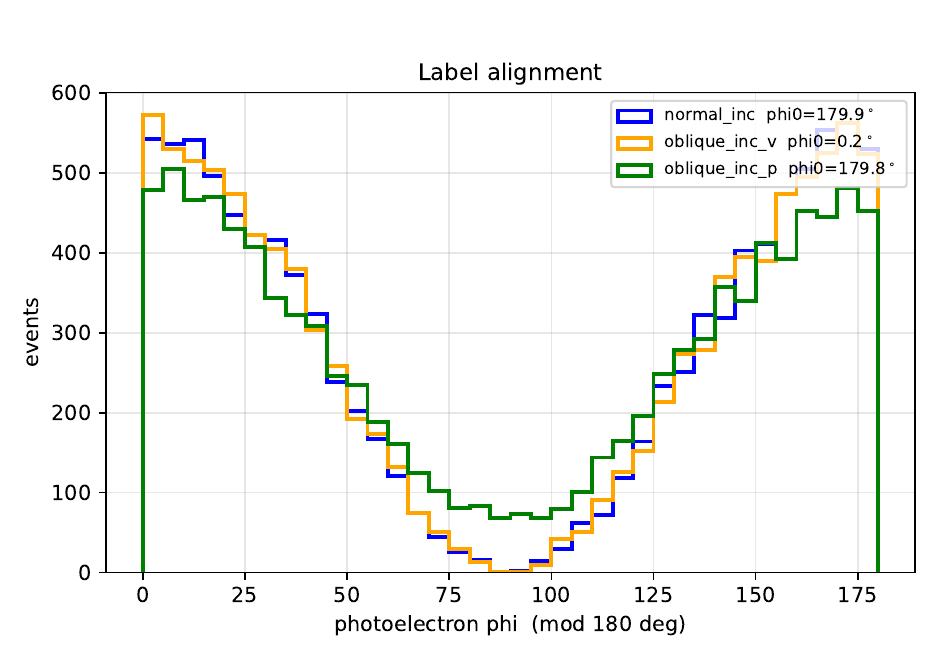}
	\caption{The distributions of labels (emission angles of photoelectrons) in simulation data for three configurations.}\label{fig:sim-label}
\end{figure}

\begin{table*}[htbp]
	\centering
	\caption{Mean absolute error (MAE) of different training schemes and architectures.}\label{tab:pred-mae}%
	\begin{tabular*}{\textwidth}{@{\extracolsep\fill}lcccc}
		\toprule
		& \multicolumn{4}{c}{MAE (degree)} \\\cmidrule{2-5}
		Method & \emph{normal\_inc} & \emph{oblique\_inc\_v} & \emph{oblique\_inc\_p} & overall \\
		\midrule
		Supervised branch-s & 17.13 & 17.21 & 21.74 & 18.69 \\
		Supervised branch-l & 15.13 & 15.76 & 20.03 & 16.98 \\
		\hline
		branch-s feat. with linear head & 18.57 & 18.44 & 23.37 & 20.13 \\
		branch-s feat. with nonlinear head & 18.40 & 18.13 & 22.99 & 19.84 \\
		branch-l feat. with linear head & 17.97 & 21.49 & 25.84 & 21.77 \\
		branch-l feat. with nonlinear head & 18.08 & 18.65 & 23.23 & 19.99 \\
		\botrule
	\end{tabular*}
\end{table*}

In the subsections above in Sect.~\ref{sec:res}, we mainly discuss the discrimination ability at the distribution level, i.e. the collective characteristics on a batch of examples. Conventionally, the X-ray polarization analysis is based on a two-stage process, in which the azimuthal emission angle of the photoelectron is predicted first at the event level. In order to apply supervised machine learning techniques, an accurate label is required for each event. We use simulation data to investigate the predictive performance of the proposed method principally because accurately labelling each event is unpractical with the detector settings.

Simulation is performed with the Star-XP software \cite{YI2024101626}. Two-dimensional images are generated along with the ground-truth azimuthal emission angles of photoelectrons as their labels, for each of the three configurations. Then, similar data preprocessing (center-of-mass aligning, cropping and normalization) transforms the generated data to the size of 40$\times$40. In Fig.~\ref{fig:sim-label}, we plot the distributions of labels for three configurations discussed above. Labels are wrapped by 180 degrees. The averages of the azimuthal angles ($\phi_0$) all locate near 0 degree or 180 degrees, with an approximate $\cos^2 \phi$ modulation. We generate 10000 examples for each configuration and use a stratified sampling strategy to randomly select 8500 examples for training and 1500 examples for testing.

The results are gathered in Table~\ref{tab:pred-mae}. For supervised branch-s and branch-l (first two rows), we change the output dimension to two so that the per-event cosine absolute loss function can be applied:
\begin{equation}
	\mathcal{L} = 1 - \left| \frac{\hat{y}_1 \cdot y_1 + \hat{y}_2 \cdot y_2}{\sqrt{\hat{y}_1^2 + \hat{y}_2^2}} \right|
\end{equation}

\noindent where $\hat{y}_1,\ \hat{y}_2$ are unnormalized network predictions, and $y_1,\ y_2$ are normalized labels (two components of the unit vector along the azimuthal angle). The above loss function can handle the discontinuity at 0/360 degrees and is insensitive to 180-degree inversion. For branch-s and branch-l features with linear or nonlinear heads (last four rows), a 64-dimensional feature vector is extracted using the same randomized neural network as discussed in Sect.~\ref{sec:main-res}. A simple head consisting of linear projection or one-hidden-layer nonlinear projection is trained on the feature vector to produce a two-dimensional output.

From Table~\ref{tab:pred-mae}, it can be seen that supervised approaches achieve lower MAE than approaches with linear/nonlinear heads on network features. For three configurations, \emph{oblique\_inc\_p} gives higher MAE than \emph{normal\_inc} and \emph{oblique\_inc\_v}. In addition, the branch-s/branch-l with the nonlinear head performs better than the same branch with the linear head. Besides, we do not observe significant performance improvements when combining the features from branch-s and branch-l.

We present a brief discussion on the above results. Since both branch-s and branch-l networks are very lightweight, the well-known issues associated with supervised training of huge network architectures (e.g. ResNet in SimCLRv2 \cite{DBLP:conf/nips/ChenKSNH20}) do not act as limiting factors for the emission angle regression task. As a result, supervised training on input images drives the weights of the entire network to converge towards optimal values, further reducing the prediction errors. However, the training process extracts the information about the regression task as much as possible and may discard other meaningful information, such as the spatially anisotropic features of the photoelectrons. A comparison of Fig.~\ref{fig:rot} and Fig.~\ref{fig:sim-label}, along with the discussion in Sect.~\ref{sec:stat}, reveals these differences.

Hence, the implication for the SGSW distance proposed in this paper is twofold. For one thing, both branch-s and branch-l network architectures are good feature extractors on the polarized images; for another, while supervised training of the network architectures on input images can yield better predictive performance, projections with random network weights can provide more comprehensive features about the original incident events.

\section{Statistical model}
\label{sec:stat}

\begin{algorithm*}
	\caption{Calculate the circular Wasserstein distance with the statistical model}\label{algo:stat}
	\begin{algorithmic}[1]
		\Require $\kappa$: kappa coefficient for the von Mises distribution; $B$: batch size.
		\Ensure $\kappa > 0$. 
		
		\Function{SingleSample}{$\theta,\ \phi,\ \gamma$}
		\State $\bm{V} \Leftarrow (\sin(\theta)\cos(\phi),\ \sin(\theta)\sin(\phi),\ \cos(
		\theta))^\top$
		\State Rotate $\bm{V}$ around rotational axis $(0,\ 1,\ 0)^\top$ by angle $\gamma$
		\State \Return $\bm{V}$
		\EndFunction
		\State
		\Function{BatchSample}{$\theta_i,\ \phi_i,\ \alpha_i$}
		\State $\Omega_\beta \Leftarrow \emptyset,\ n \Leftarrow 0 $
		\While{$n < B$}
		\State Sample $\theta_o,\ \phi_o \sim p(\theta,\ \phi) = \frac{\sin^2(\theta)\sin^2(\phi)}{\pi^2},\ \theta,\ \phi\in\left[0,\ 2\pi\right)$
		\State $\bm{V}_n \Leftarrow \textsc{SingleSample}(\theta_o,\ \phi_o + \phi_i,\ \theta_i)$
		\State $\alpha_o \Leftarrow \arctan 2(V_{n,y}, V_{n,x})$
		\State Sample $\beta \sim \textsc{vonMises}(\kappa,\ \alpha_o + \alpha_i)$
		\State $\Omega_\beta \Leftarrow \Omega_\beta \bigcup \{ \beta \}$
		\State $n \Leftarrow n + 1$
		\EndWhile
		\State \Return $\Omega_\beta$
		\EndFunction
		\State
		\Procedure{ComputeCW}{$\theta_{i,1},\ \phi_{i,1},\ \alpha_{i,1},\ \theta_{i,2},\ \phi_{i,2},\ \alpha_{i,2}$}
		\State $\Omega_{\beta,1} \Leftarrow \textsc{BatchSample}(\theta_{i,1},\ \phi_{i,1},\ \alpha_{i,1})$
		\State $\Omega_{\beta,2} \Leftarrow \textsc{BatchSample}(\theta_{i,2},\ \phi_{i,2},\ \alpha_{i,2})$
		\State $W \Leftarrow \textsc{CircularWasserstein}(\Omega_{\beta,1},\ \Omega_{\beta,2})$
		\EndProcedure
	\end{algorithmic}
\end{algorithm*}

\begin{figure*}[htbp]
	\centering
	\includegraphics[width=0.9\textwidth]{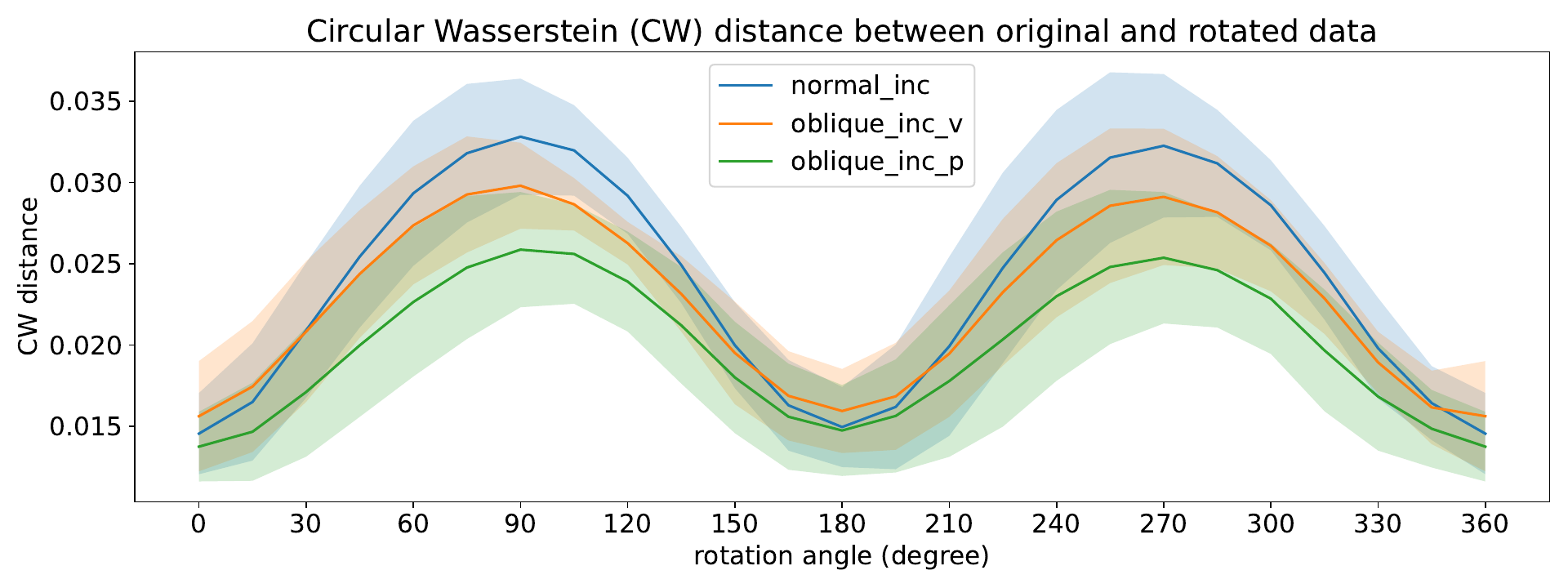}
	\caption{The curves of the circular Wasserstein distances on the statistical model between the distribution of original angular samples and the distributions of angular samples after rotation.}\label{fig:stat}
\end{figure*}

\begin{table*}[htbp]
	\caption{Comparison on the peak-to-valley ratio and the average standard deviation (std.) to mean between the curves of the SGSW distances (Experimental) and the CW distances (Statistical Model). For experimental data, we manipulate the polarized images from the detector. For statistical modelling, we use different input parameters for calculation. The values on the curves in Fig.~\ref{fig:rot} and Fig.~\ref{fig:stat} are obtained at a series of rotation angles.}\label{tab:stat}
	\begin{tabular*}{\textwidth}{@{\extracolsep\fill}lcccc}
		\toprule%
		& \multicolumn{2}{@{}c@{}}{Peak-to-valley ratio} & \multicolumn{2}{@{}c@{}}{Average std. to mean} \\\cmidrule{2-3}\cmidrule{4-5}%
		Configuration & Experimental & Statistical Model & Experimental & Statistical Model \\
		\midrule
		\emph{normal\_inc}     & 2.2352 & 2.2548 & 0.1469 & 0.1566\\
		\emph{oblique\_inc\_v} & 1.8267 & 1.9073 & 0.1456 & 0.1485\\
		\emph{oblique\_inc\_p} & 1.5747 & 1.8810 & 0.1519 & 0.1863\\
		\botrule
	\end{tabular*}
\end{table*}

This section presents a simplified statistical model to demonstrate the physical significance of the numerical results obtained by rotating the polarized images. The proposed SGSW distance (in this case, the GSW distance of branch-l) is from numerical computation without any assumption on physical principles. Although the discrimination ability is what we care about most, we also hope this metric, at its relative value, could match the physical process of the detector. The statistical model is invented with the purpose of showing the SGSW distance does carry physical information, instead of purely working as discriminators.

In the statistical model, apart from geometric transformations, we make one core assumption that the detected emission angle of the photoelectron obeys the von Mises distribution \cite{von1964mathematical} centering the real emission angle. Under this assumption, we can calculate the circular Wasserstein (CW) distance \cite{Hundrieser2022} between two angular distributions. The choices of the von Mises distribution and the CW distance are based on the fact that we perform statistics on a circular plane.

The algorithmic procedure of computing the CW distance on the statistical model is shown in Algorithm \ref{algo:stat}. In this algorithm, the function \textsc{SingleSample} produces a unit vector in the space rectangular coordinate system from polar and azimuthal angles, and performs an additional rotation around the y-axis. The function \textsc{BatchSample} produces a collection of detected emission angles. Firstly, the polar and azimuthal angles are sampled from a squared-sine distribution according to equation (\ref{equ:angle}) (the relativistic effect is not considered for simplicity, and the default direction of the electric vector is along the y-axis). Then we sample a unit vector using the aforementioned function with additional rotation and adjustment to the azimuthal angle to simulate different configurations. Afterwards, the direction on the x-y plane is determined by the unit vector, and an angle is sampled from the von Mises distribution, with a predefined kappa coefficient, centering the direction on the x-y plane. Finally, this sampled angle is added to the collection. We repeat the sampling loop until we get enough examples to form a sufficient angular distribution.

When computing the CW distance, we follow the procedure \textsc{ComputeCW}. Two batches are sampled with independent configurations and rotations. The $\theta_{i,*},\ \phi_{i,*},\ \alpha_{i,*}$ correspond to the angle of inclination for oblique incidence, the direction of the electric vector, and the rotation angle, respectively. Then the circular Wasserstein distance is computed on the two batches.

When implementing the algorithm, we use SciPy \cite{2020SciPy-NMeth} and POT \cite{flamary2021pot} libraries in Python for the von Mises distribution sampling and CW distance computation. We use the batch size of 512, and compute the mean and the standard deviation on 64 batches. The results for three configurations on different rotation angles are shown in Fig.~\ref{fig:stat}. The only tunable parameter for the algorithm is the kappa coefficient for the von Mises distribution. We manually adjust kappa (which is set to 1.85 in the model) to match the amplitudes proportionally between the experimental results and the statistical model. By tuning the parameter, we try to make the peak value and valley value of the statistical model have the same ratio as in the experiment. Nevertheless, inspection on a single case would not be meaningful. What we care about here is the inter-relation between three different configurations (\emph{normal\_inc}, \emph{oblique\_inc\_v} and \emph{oblique\_inc\_p}). It would be probable that the SGSW distance conforms to certain physical principles of the detector if all configurations are matched between the experimental results and the statistical model.

By comparing Fig.~\ref{fig:stat} and Fig.~\ref{fig:rot}, we can find that they share many similarities. The shapes of the curves are very alike, the peaks for three configurations (\emph{normal\_inc}, \emph{oblique\_inc\_v}, \emph{oblique\_inc\_p}) have the same ordering, and the shades indicating the standard deviations are on the same scale.

To numerically compare the two figures, we calculate the peak-to-valley ratio and the average standard deviation to mean for three configurations with two kinds of Wasserstein distances. The results are shown in Table~\ref{tab:stat}. It can be seen that the two separate analysis procedures have very similar quantities. Among the configurations, \emph{normal\_inc} and \emph{oblique\_inc\_v} can match very closely, while the statistical model slightly overestimates the quantities for \emph{oblique\_inc\_p}. In summary, it is demonstrated that a statistical model with simple assumptions can very well fit the experimental results. 

\section{Conclusions and outlook}
\label{sec:con}

In this paper, we present a data-driven approach for GPD-based polarization analysis on direct charge images in astroparticle experiments. A prototype of the LPD in the POLAR-2 experiment is constructed for researching soft X-rays on the keV scale. Images from three configurations with varied incident angles and polarization directions are collected on the platform of the detector prototype. Initial observation reveals the subtle distinctions between the image patterns from different configurations.

The major contribution of the paper is a method to compare the data distributions of two image sets through the SGSW distance, using an ensemble of neural networks with randomly initialized weights. This method is applied to analyzing the polarized images from GPD to reveal the mismatches in angular distributions of photoelectrons. Unlike conventional supervised or unsupervised/self-supervised approaches, the proposed method does not depend on explicit training or optimization, making it a good candidate for raw experimental data without labelling by extra equipments or modelling in simulation.

In experiments, we evaluate the SGSW distance on image data from several configurations and also azimuthal rotations. The results show that the proposed distance measure has the ability to distinguish between different image patterns and angular distributions. What we find empirically is that the dual-branch structures of the neural network work in a complementary manner, with each branch focusing on distinct aspects of the data distribution. These results can be a first step towards discriminative analysis with more sophisticated neural network architectures, paving the way for future researches.

Finally, we build a simplified statistical model to demonstrate the validity of the experimental results. The statistical model is based on the von Mises distribution and the CW distance. We give an algorithmic procedure to implement the statistical model and to compute the CW distance systematically. The results from the statistical model show high consistency with results from the experiments, which cross-validates the credibility from physical and statistical perspectives.

The method and results reported here have extensive implications. For GPD-based astroparticle experiments, this method can be an effective procedure to assist polarization analysis prior to conventional feature extraction. In particular, for POLAR-2/LPD, the wide field of view necessitates analytical measures to acquire additional information from incident signal events and to facilitate precise measurement of X-ray polarization.

Apart from applications in astroparticle physics, the method can also be employed in any circumstances where there are abundant image data. For example, the SGSW distance can be used for anomaly detection and searching for new physics in high energy physics experiments \cite{Dort2022}. Besides, it can be used in generative models to evaluate the quality of generation and to mitigate the gap between simulation to reality \cite{Erdmann2023,Fu2024}. It is worth mentioning that the applications are not limited to 2D spaces. As a matter of course, it works naturally with one-dimensional or three-dimensional data.

Future work includes investigating the proposed method in more application scenarios, enhancing the discrimination power with dedicated neural network architectures, and improving the robustness of the method when it is used with fewer examples and more complex data distributions. We sincerely hope the method proposed here will benefit researchers and practitioners in this area as an effective computational tool for more real-world experiments.

\backmatter

\bmhead{Funding}

This research is supported by the National Natural Science Foundation of China (Grant Nos. 12405228, U1731239, 12027803, 12133003), the National Key Research and Development Program of China (Grant Nos. 2024YFF0726201, 2024YFA1611700), the China Postdoctoral Science Foundation (Grant No. 2023M731244), the special funding for Guangxi Bagui Scholars, the Guangxi Key Research and Development Program (Guike FN2504240040), the Guangxi Talent Program (``Highland of Innovation Talents''), the Innovation Project of Guangxi Graduate Education (No. YCBZ2025045), and the Fundamental Research Funds for the Central Universities (Grant No. CCNU25PTFZ013).

\bmhead{Data Availability Statement}

This data will be made available on reasonable request. [Authors' comment: The data can be made available upon reasonable request from the author Huanbo Feng.]

\bmhead{Code Availability Statement}

Code/software will be made available on reasonable request. [Authors' comment: The code/software generated and/or analysed during the current study can be made available upon reasonable request from the corresponding author.]

\appendix

\section{Details of network architectures and settings}
\label{sec:detail-arch}

\subsection{Details of self-supervised models}
\label{sec:detail-self-sup}

\paragraph{SimCLRv2.} We use the pre-training mode for encoder network optimization. The encoder network is an 18-layer ResNet \cite{7780459}, mapping the input image (40$\times$40$\times$1) to a 512-length feature vector. For data augmentation, we drop cropping and flipping: since the image is always aligned to the center of mass, no cropping is needed; besides, since the generated feature vector should be sensitive to the rotation of the image, no flipping is used. The main augmentation includes blurring and variations of brightness, contrast, saturation and hue. We train the network on the entire datasets of polarized images for 50 epochs with the batch size of 128 on a single graphics processing unit (GPU). Other training parameters are the same as those in the SimCLRv2 repository.

\paragraph{DINOv2.} We create a tiny ViT \cite{DBLP:conf/iclr/DosovitskiyB0WZ21} network for the image feature extraction. The network contains 6 stacked transformer blocks, with 3 attention heads in each self-attention layer, and a 4:1 ratio of the hidden size to the embedding dimension for the multi-layer perceptron. A 192-length feature vector is mapped from the input image (40$\times$40$\times$1) by the network. We use data augmentation techniques similar to those used for SimCLRv2 discussed above. We train the network on the entire datasets of polarized images for 50 epochs (with 5 warm-up epochs) with the batch size of 64 on a single GPU. Other training parameters are the same as those in the DINOv2 repository.

\subsection{Details of models for architectural research}
\label{sec:detail-sens-arch}

The networks are initialized with the same method discussed in Sect.~\ref{sec:method}. Each one is used as individual networks for the GSW distance.

\paragraph{LeNet.} We create a LeNet-5 variant as a sequential network. The network is composed of: (1) 2D convolution: 6 kernels, size of 5, unit stride, with tanh activation; (2) 2D max pooling: size of 2, stride of 2; (3) 2D convolution: 16 kernels, size of 5, unit stride, with tanh activation; (4) 2D max pooling: size of 2, stride of 2; (5) flattening; (6) dense connection: 120 hidden units, with tanh activation; (7) dense connection: 1 output unit, with no activation.

\paragraph{MobileNetV2.} We create a simplified MobileNetV2 variant with a sequential structure. The width multiplier is set to 1.0. The network is composed of: (1) 3$\times$3 convolution followed by batch normalization: output channel of 16, with ReLU6 activation; (2) 5 successive inverted residual blocks with parameters reduced for adaptation: expanding ratio of (1, 4, 4, 4, 4), channel factor of (16, 24, 32, 64, 96), layer repetition of (1, 2, 2, 2, 1), and stride of (1, 2, 2, 2, 1); (3) 1$\times$1 convolution followed by batch normalization: output channel of 256, with ReLU6 activation; (4) adaptive average pooling: unit size; (5) flattening; (6) dense connection: 1 output unit, with no activation.

\paragraph{EfficientNet.} We create a simplified EfficientNet variant with a sequential structure. The network is composed of: (1) 3$\times$3 convolution followed by batch normalization: output channel of 16, with SiLU activation; (2) 4 successive mobile inverted bottleneck-style blocks: output channel of (16, 24, 32, 48), expanding ratio of (1, 2, 2, 2), stride of 2; (3) 1$\times$1 convolution followed by batch normalization: output channel of 64, with SiLU activation; (4) adaptive average pooling: unit size; (5) flattening; (6) dense connection: 1 output unit, with no activation.


\bibliography{mybibfile}


\begin{thebibliography}{46}
\ifx \bisbn   \undefined \def \bisbn  #1{ISBN #1}\fi
\ifx \binits  \undefined \def \binits#1{#1}\fi
\ifx \bauthor  \undefined \def \bauthor#1{#1}\fi
\ifx \batitle  \undefined \def \batitle#1{#1}\fi
\ifx \bjtitle  \undefined \def \bjtitle#1{#1}\fi
\ifx \bvolume  \undefined \def \bvolume#1{\textbf{#1}}\fi
\ifx \byear  \undefined \def \byear#1{#1}\fi
\ifx \bissue  \undefined \def \bissue#1{#1}\fi
\ifx \bfpage  \undefined \def \bfpage#1{#1}\fi
\ifx \blpage  \undefined \def \blpage #1{#1}\fi
\ifx \burl  \undefined \def \burl#1{\textsf{#1}}\fi
\ifx \doiurl  \undefined \def \doiurl#1{\url{https://doi.org/#1}}\fi
\ifx \betal  \undefined \def \betal{\textit{et al.}}\fi
\ifx \binstitute  \undefined \def \binstitute#1{#1}\fi
\ifx \binstitutionaled  \undefined \def \binstitutionaled#1{#1}\fi
\ifx \bctitle  \undefined \def \bctitle#1{#1}\fi
\ifx \beditor  \undefined \def \beditor#1{#1}\fi
\ifx \bpublisher  \undefined \def \bpublisher#1{#1}\fi
\ifx \bbtitle  \undefined \def \bbtitle#1{#1}\fi
\ifx \bedition  \undefined \def \bedition#1{#1}\fi
\ifx \bseriesno  \undefined \def \bseriesno#1{#1}\fi
\ifx \blocation  \undefined \def \blocation#1{#1}\fi
\ifx \bsertitle  \undefined \def \bsertitle#1{#1}\fi
\ifx \bsnm \undefined \def \bsnm#1{#1}\fi
\ifx \bsuffix \undefined \def \bsuffix#1{#1}\fi
\ifx \bparticle \undefined \def \bparticle#1{#1}\fi
\ifx \barticle \undefined \def \barticle#1{#1}\fi
\bibcommenthead
\ifx \bconfdate \undefined \def \bconfdate #1{#1}\fi
\ifx \botherref \undefined \def \botherref #1{#1}\fi
\ifx \url \undefined \def \url#1{\textsf{#1}}\fi
\ifx \bchapter \undefined \def \bchapter#1{#1}\fi
\ifx \bbook \undefined \def \bbook#1{#1}\fi
\ifx \bcomment \undefined \def \bcomment#1{#1}\fi
\ifx \oauthor \undefined \def \oauthor#1{#1}\fi
\ifx \citeauthoryear \undefined \def \citeauthoryear#1{#1}\fi
\ifx \endbibitem  \undefined \def \endbibitem {}\fi
\ifx \bconflocation  \undefined \def \bconflocation#1{#1}\fi
\ifx \arxivurl  \undefined \def \arxivurl#1{\textsf{#1}}\fi
\csname PreBibitemsHook\endcsname

\bibitem[\protect\citeauthoryear{Wang et~al.}{2024}]{Wang_2024}
\begin{barticle}
\bauthor{\bsnm{Wang}, \binits{J.}},
\bauthor{\bsnm{Wang}, \binits{H.}},
\bauthor{\bsnm{Zhang}, \binits{D.}}:
\batitle{{Solar neutrino background in high-pressure gaseous 82SeF6 TPC
  neutrinoless double beta decay experiments}}.
\bjtitle{Chinese Physics C}
\bvolume{48}(\bissue{4}),
\bfpage{043003}
(\byear{2024})
\doiurl{10.1088/1674-1137/ad2675}
\end{barticle}
\endbibitem

\bibitem[\protect\citeauthoryear{Filipenko et~al.}{2013}]{Filipenko2013}
\begin{barticle}
\bauthor{\bsnm{Filipenko}, \binits{M.}},
\bauthor{\bsnm{Gleixner}, \binits{T.}},
\bauthor{\bsnm{Anton}, \binits{G.}},
\bauthor{\bsnm{Durst}, \binits{J.}},
\bauthor{\bsnm{Michel}, \binits{T.}}:
\batitle{{Characterization of the energy resolution and the tracking
  capabilities of a hybrid pixel detector with CdTe-sensor layer for a possible
  use in a neutrinoless double beta decay experiment}}.
\bjtitle{The European Physical Journal C}
\bvolume{73}(\bissue{4}),
\bfpage{2374}
(\byear{2013})
\doiurl{10.1140/epjc/s10052-013-2374-1}
\end{barticle}
\endbibitem

\bibitem[\protect\citeauthoryear{Amaro et~al.}{2022}]{instruments6010006}
\begin{botherref}
\oauthor{\bsnm{Amaro}, \binits{F.D.}}, et al.:
{The CYGNO Experiment}.
Instruments
\textbf{6}(1)
(2022)
\doiurl{10.3390/instruments6010006}
\end{botherref}
\endbibitem

\bibitem[\protect\citeauthoryear{Yi et~al.}{2026}]{Yi2026}
\begin{barticle}
\bauthor{\bsnm{Yi}, \binits{D.}}, \betal:
\batitle{{Direct observation of the Migdal effect induced by neutron
  bombardment}}.
\bjtitle{Nature}
\bvolume{649}(\bissue{8097}),
\bfpage{580}--\blpage{583}
(\byear{2026})
\doiurl{10.1038/s41586-025-09918-8}
\end{barticle}
\endbibitem

\bibitem[\protect\citeauthoryear{Wang et~al.}{2022}]{Wang2022}
\begin{barticle}
\bauthor{\bsnm{Wang}, \binits{H.-L.}}, \betal:
\batitle{{Design and tests of the prototype beam monitor of the CSR external
  target experiment}}.
\bjtitle{Nuclear Science and Techniques}
\bvolume{33}(\bissue{3}),
\bfpage{36}
(\byear{2022})
\doiurl{10.1007/s41365-022-01021-1}
\end{barticle}
\endbibitem

\bibitem[\protect\citeauthoryear{Weisskopf et~al.}{2016}]{WEISSKOPF20161179}
\begin{barticle}
\bauthor{\bsnm{Weisskopf}, \binits{M.C.}}, \betal:
\batitle{{The Imaging X-ray Polarimetry Explorer (IXPE)}}.
\bjtitle{Results in Physics}
\bvolume{6},
\bfpage{1179}--\blpage{1180}
(\byear{2016})
\doiurl{10.1016/j.rinp.2016.10.021}
\end{barticle}
\endbibitem

\bibitem[\protect\citeauthoryear{Feng et~al.}{2019}]{Feng2019}
\begin{barticle}
\bauthor{\bsnm{Feng}, \binits{H.}}, \betal:
\batitle{{PolarLight: a CubeSat X-ray polarimeter based on the gas pixel
  detector}}.
\bjtitle{Experimental Astronomy}
\bvolume{47}(\bissue{1}),
\bfpage{225}--\blpage{243}
(\byear{2019})
\doiurl{10.1007/s10686-019-09625-z}
\end{barticle}
\endbibitem

\bibitem[\protect\citeauthoryear{Feng et~al.}{2024}]{Feng2024}
\begin{barticle}
\bauthor{\bsnm{Feng}, \binits{H.-B.}}, \betal:
\batitle{{Gas microchannel plate-pixel detector for X-ray polarimetry}}.
\bjtitle{Nuclear Science and Techniques}
\bvolume{35}(\bissue{5}),
\bfpage{39}
(\byear{2024})
\doiurl{10.1007/s41365-024-01407-3}
\end{barticle}
\endbibitem

\bibitem[\protect\citeauthoryear{Ai et~al.}{2018}]{Ai_2018}
\begin{barticle}
\bauthor{\bsnm{Ai}, \binits{P.}},
\bauthor{\bsnm{Wang}, \binits{D.}},
\bauthor{\bsnm{Huang}, \binits{G.}},
\bauthor{\bsnm{Sun}, \binits{X.}}:
\batitle{{Three-dimensional convolutional neural networks for neutrinoless
  double-beta decay signal/background discrimination in high-pressure gaseous
  Time Projection Chamber}}.
\bjtitle{Journal of Instrumentation}
\bvolume{13}(\bissue{08}),
\bfpage{08015}
(\byear{2018})
\doiurl{10.1088/1748-0221/13/08/P08015}
\end{barticle}
\endbibitem

\bibitem[\protect\citeauthoryear{Amaro et~al.}{2025}]{Amaro2025}
\begin{barticle}
\bauthor{\bsnm{Amaro}, \binits{F.D.}}, \betal:
\batitle{{Bayesian network 3D event reconstruction in the Cygno optical TPC for
  dark matter direct detection}}.
\bjtitle{The European Physical Journal C}
\bvolume{85}(\bissue{11}),
\bfpage{1261}
(\byear{2025})
\doiurl{10.1140/epjc/s10052-025-14965-6}
\end{barticle}
\endbibitem

\bibitem[\protect\citeauthoryear{Ai et~al.}{2020}]{AI2020164640}
\begin{barticle}
\bauthor{\bsnm{Ai}, \binits{P.}},
\bauthor{\bsnm{Wang}, \binits{D.}},
\bauthor{\bsnm{Sun}, \binits{X.}},
\bauthor{\bsnm{Huang}, \binits{G.}},
\bauthor{\bsnm{Li}, \binits{Z.}}:
\batitle{{A deep learning approach to multi-track location and orientation in
  gaseous drift chambers}}.
\bjtitle{Nuclear Instruments and Methods in Physics Research Section A:
  Accelerators, Spectrometers, Detectors and Associated Equipment}
\bvolume{984},
\bfpage{164640}
(\byear{2020})
\doiurl{10.1016/j.nima.2020.164640}
\end{barticle}
\endbibitem

\bibitem[\protect\citeauthoryear{Jiao et~al.}{2025}]{Jiao2025}
\begin{barticle}
\bauthor{\bsnm{Jiao}, \binits{Y.}}, \betal:
\batitle{{Optimization of deep learning method on track reconstruction for
  X-ray polarimetry with gas pixel detectors}}.
\bjtitle{Experimental Astronomy}
\bvolume{59}(\bissue{3}),
\bfpage{34}
(\byear{2025})
\doiurl{10.1007/s10686-025-10003-1}
\end{barticle}
\endbibitem

\bibitem[\protect\citeauthoryear{Detlefs et~al.}{2012}]{Detlefs2012}
\begin{barticle}
\bauthor{\bsnm{Detlefs}, \binits{C.}},
\bauthor{\bsnm{Rio}, \binits{M.}},
\bauthor{\bsnm{Mazzoli}, \binits{C.}}:
\batitle{{X-ray polarization: General formalism and polarization analysis}}.
\bjtitle{The European Physical Journal Special Topics}
\bvolume{208}(\bissue{1}),
\bfpage{359}--\blpage{371}
(\byear{2012})
\doiurl{10.1140/epjst/e2012-01630-3}
\end{barticle}
\endbibitem

\bibitem[\protect\citeauthoryear{Huang et~al.}{2021}]{Huang2021}
\begin{barticle}
\bauthor{\bsnm{Huang}, \binits{X.-F.}}, \betal:
\batitle{{Simulation and photoelectron track reconstruction of soft X-ray
  polarimeter}}.
\bjtitle{Nuclear Science and Techniques}
\bvolume{32}(\bissue{7}),
\bfpage{67}
(\byear{2021})
\doiurl{10.1007/s41365-021-00903-0}
\end{barticle}
\endbibitem

\bibitem[\protect\citeauthoryear{Puetter and Yahil}{1999}]{Puetter:1999qb}
\begin{barticle}
\bauthor{\bsnm{Puetter}, \binits{R.C.}},
\bauthor{\bsnm{Yahil}, \binits{A.}}:
\batitle{{The pixon method of image reconstruction}}.
\bjtitle{ASP Conf. Ser.}
\bvolume{172},
\bfpage{307}
(\byear{1999})
{\href{https://arxiv.org/abs/astro-ph/9901063}{{arXiv:astro-ph/9901063}}}
\end{barticle}
\endbibitem

\bibitem[\protect\citeauthoryear{Panaretos and Zemel}{2019}]{Panaretos2019}
\begin{barticle}
\bauthor{\bsnm{Panaretos}, \binits{V.M.}},
\bauthor{\bsnm{Zemel}, \binits{Y.}}:
\batitle{{Statistical Aspects of Wasserstein Distances}}.
\bjtitle{Annual Review of Statistics and Its Application}
\bvolume{6},
\bfpage{405}--\blpage{431}
(\byear{2019})
\doiurl{10.1146/annurev-statistics-030718-104938}
\end{barticle}
\endbibitem

\bibitem[\protect\citeauthoryear{Kolouri et~al.}{2018}]{8578459}
\begin{bchapter}
\bauthor{\bsnm{Kolouri}, \binits{S.}},
\bauthor{\bsnm{Rohde}, \binits{G.K.}},
\bauthor{\bsnm{Hoffmann}, \binits{H.}}:
\bctitle{{Sliced Wasserstein Distance for Learning Gaussian Mixture Models}}.
In: \bbtitle{2018 IEEE/CVF Conference on Computer Vision and Pattern
  Recognition},
pp. \bfpage{3427}--\blpage{3436}
(\byear{2018}).
\doiurl{10.1109/CVPR.2018.00361}
\end{bchapter}
\endbibitem

\bibitem[\protect\citeauthoryear{Kolouri
  et~al.}{2019}]{DBLP:conf/nips/KolouriNSBR19}
\begin{bchapter}
\bauthor{\bsnm{Kolouri}, \binits{S.}},
\bauthor{\bsnm{Nadjahi}, \binits{K.}},
\bauthor{\bsnm{Simsekli}, \binits{U.}},
\bauthor{\bsnm{Badeau}, \binits{R.}},
\bauthor{\bsnm{Rohde}, \binits{G.K.}}:
\bctitle{{Generalized Sliced Wasserstein Distances}}.
In: \bbtitle{Advances in Neural Information Processing Systems 32: Annual
  Conference on Neural Information Processing Systems 2019, NeurIPS 2019,
  December 8-14, 2019, Vancouver, BC, Canada},
pp. \bfpage{261}--\blpage{272}
(\byear{2019})
\end{bchapter}
\endbibitem

\bibitem[\protect\citeauthoryear{Nietert
  et~al.}{2022}]{DBLP:conf/nips/NietertGSK22}
\begin{bchapter}
\bauthor{\bsnm{Nietert}, \binits{S.}},
\bauthor{\bsnm{Goldfeld}, \binits{Z.}},
\bauthor{\bsnm{Sadhu}, \binits{R.}},
\bauthor{\bsnm{Kato}, \binits{K.}}:
\bctitle{{Statistical, Robustness, and Computational Guarantees for Sliced
  Wasserstein Distances}}.
In: \bbtitle{Advances in Neural Information Processing Systems 35: Annual
  Conference on Neural Information Processing Systems 2022, NeurIPS 2022, New
  Orleans, LA, USA, November 28 - December 9, 2022}
(\byear{2022})
\end{bchapter}
\endbibitem

\bibitem[\protect\citeauthoryear{Deshpande et~al.}{2018}]{8578465}
\begin{bchapter}
\bauthor{\bsnm{Deshpande}, \binits{I.}},
\bauthor{\bsnm{Zhang}, \binits{Z.}},
\bauthor{\bsnm{Schwing}, \binits{A.}}:
\bctitle{{Generative Modeling Using the Sliced Wasserstein Distance}}.
In: \bbtitle{2018 IEEE/CVF Conference on Computer Vision and Pattern
  Recognition},
pp. \bfpage{3483}--\blpage{3491}
(\byear{2018}).
\doiurl{10.1109/CVPR.2018.00367}
\end{bchapter}
\endbibitem

\bibitem[\protect\citeauthoryear{Kitaguchi et~al.}{2019}]{KITAGUCHI2019162389}
\begin{barticle}
\bauthor{\bsnm{Kitaguchi}, \binits{T.}}, \betal:
\batitle{{A convolutional neural network approach for reconstructing
  polarization information of photoelectric X-ray polarimeters}}.
\bjtitle{Nuclear Instruments and Methods in Physics Research Section A:
  Accelerators, Spectrometers, Detectors and Associated Equipment}
\bvolume{942},
\bfpage{162389}
(\byear{2019})
\doiurl{10.1016/j.nima.2019.162389}
\end{barticle}
\endbibitem

\bibitem[\protect\citeauthoryear{Li et~al.}{2025}]{Li2025}
\begin{barticle}
\bauthor{\bsnm{Li}, \binits{Y.-N.}}, \betal:
\batitle{{Research on the X-ray polarization deconstruction method based on
  hexagonal convolutional neural network}}.
\bjtitle{Nuclear Science and Techniques}
\bvolume{36}(\bissue{2}),
\bfpage{31}
(\byear{2025})
\doiurl{10.1007/s41365-024-01598-9}
\end{barticle}
\endbibitem

\bibitem[\protect\citeauthoryear{Moriakov
  et~al.}{2020}]{DBLP:journals/corr/abs-2005-08126}
\begin{botherref}
\oauthor{\bsnm{Moriakov}, \binits{N.}},
\oauthor{\bsnm{Samudre}, \binits{A.}},
\oauthor{\bsnm{Negro}, \binits{M.}},
\oauthor{\bsnm{Gieseke}, \binits{F.}},
\oauthor{\bsnm{Otten}, \binits{S.}},
\oauthor{\bsnm{Hendriks}, \binits{L.}}:
{Inferring astrophysical X-ray polarization with deep learning}.
CoRR
\textbf{abs/2005.08126}
(2020)
{\href{https://arxiv.org/abs/2005.08126}{{2005.08126}}}
\end{botherref}
\endbibitem

\bibitem[\protect\citeauthoryear{Dort et~al.}{2022}]{Dort2022}
\begin{barticle}
\bauthor{\bsnm{Dort}, \binits{K.}},
\bauthor{\bsnm{Bilk}, \binits{J.}},
\bauthor{\bsnm{Käs}, \binits{S.}},
\bauthor{\bsnm{Lange}, \binits{J.S.}},
\bauthor{\bsnm{Peter}, \binits{M.}},
\bauthor{\bsnm{Schellhaas}, \binits{T.}},
\bauthor{\bsnm{Schwenker}, \binits{B.}},
\bauthor{\bsnm{Spruck}, \binits{B.}}:
\batitle{{Comparison of supervised and unsupervised anomaly detection in Belle
  II pixel detector data}}.
\bjtitle{The European Physical Journal C}
\bvolume{82}(\bissue{7}),
\bfpage{587}
(\byear{2022})
\doiurl{10.1140/epjc/s10052-022-10548-x}
\end{barticle}
\endbibitem

\bibitem[\protect\citeauthoryear{Chen et~al.}{2020}]{DBLP:conf/nips/ChenKSNH20}
\begin{bchapter}
\bauthor{\bsnm{Chen}, \binits{T.}},
\bauthor{\bsnm{Kornblith}, \binits{S.}},
\bauthor{\bsnm{Swersky}, \binits{K.}},
\bauthor{\bsnm{Norouzi}, \binits{M.}},
\bauthor{\bsnm{Hinton}, \binits{G.E.}}:
\bctitle{{Big Self-Supervised Models are Strong Semi-Supervised Learners}}.
In: \bbtitle{Advances in Neural Information Processing Systems 33: Annual
  Conference on Neural Information Processing Systems 2020, NeurIPS 2020,
  December 6-12, 2020, Virtual}
(\byear{2020})
\end{bchapter}
\endbibitem

\bibitem[\protect\citeauthoryear{Oquab
  et~al.}{2024}]{DBLP:journals/tmlr/OquabDMVSKFHMEA24}
\begin{botherref}
\oauthor{\bsnm{Oquab}, \binits{M.}}, et al.:
{DINOv2: Learning Robust Visual Features without Supervision}.
Trans. Mach. Learn. Res.
\textbf{2024}
(2024)
\end{botherref}
\endbibitem

\bibitem[\protect\citeauthoryear{Hulsman}{2020}]{10.1117/12.2559374}
\begin{bchapter}
\bauthor{\bsnm{Hulsman}, \binits{J.}}:
\bctitle{{POLAR-2: a large scale gamma-ray polarimeter for GRBs}}.
In: \bbtitle{Space Telescopes and Instrumentation 2020: Ultraviolet to Gamma
  Ray},
vol. \bseriesno{11444},
p. \bfpage{114442}
(\byear{2020}).
\doiurl{10.1117/12.2559374}
\end{bchapter}
\endbibitem

\bibitem[\protect\citeauthoryear{An et~al.}{2016}]{AN2016144}
\begin{barticle}
\bauthor{\bsnm{An}, \binits{M.}}, \betal:
\batitle{{A low-noise CMOS pixel direct charge sensor, Topmetal-II-}}.
\bjtitle{Nuclear Instruments and Methods in Physics Research Section A:
  Accelerators, Spectrometers, Detectors and Associated Equipment}
\bvolume{810},
\bfpage{144}--\blpage{150}
(\byear{2016})
\doiurl{10.1016/j.nima.2015.11.153}
\end{barticle}
\endbibitem

\bibitem[\protect\citeauthoryear{Li et~al.}{2021}]{LI2021165430}
\begin{barticle}
\bauthor{\bsnm{Li}, \binits{Z.}}, \betal:
\batitle{{Preliminary test of topmetal-II- sensor for X-ray polarization
  measurements}}.
\bjtitle{Nuclear Instruments and Methods in Physics Research Section A:
  Accelerators, Spectrometers, Detectors and Associated Equipment}
\bvolume{1008},
\bfpage{165430}
(\byear{2021})
\doiurl{10.1016/j.nima.2021.165430}
\end{barticle}
\endbibitem

\bibitem[\protect\citeauthoryear{Gavrila}{1959}]{PhysRev.113.514}
\begin{barticle}
\bauthor{\bsnm{Gavrila}, \binits{M.}}:
\batitle{{Relativistic $K$-Shell Photoeffect}}.
\bjtitle{Phys. Rev.}
\bvolume{113},
\bfpage{514}--\blpage{526}
(\byear{1959})
\doiurl{10.1103/PhysRev.113.514}
\end{barticle}
\endbibitem

\bibitem[\protect\citeauthoryear{Xie et~al.}{2023}]{Xie2023}
\begin{barticle}
\bauthor{\bsnm{Xie}, \binits{Y.}}, \betal:
\batitle{{Variably polarized X-ray sources for LPD calibration}}.
\bjtitle{Experimental Astronomy}
\bvolume{56}(\bissue{2}),
\bfpage{499}--\blpage{515}
(\byear{2023})
\doiurl{10.1007/s10686-023-09905-9}
\end{barticle}
\endbibitem

\bibitem[\protect\citeauthoryear{Li et~al.}{2022}]{9451544}
\begin{barticle}
\bauthor{\bsnm{Li}, \binits{Z.}},
\bauthor{\bsnm{Liu}, \binits{F.}},
\bauthor{\bsnm{Yang}, \binits{W.}},
\bauthor{\bsnm{Peng}, \binits{S.}},
\bauthor{\bsnm{Zhou}, \binits{J.}}:
\batitle{{A Survey of Convolutional Neural Networks: Analysis, Applications,
  and Prospects}}.
\bjtitle{IEEE Transactions on Neural Networks and Learning Systems}
\bvolume{33}(\bissue{12}),
\bfpage{6999}--\blpage{7019}
(\byear{2022})
\doiurl{10.1109/TNNLS.2021.3084827}
\end{barticle}
\endbibitem

\bibitem[\protect\citeauthoryear{He et~al.}{2015}]{DBLP:conf/iccv/HeZRS15}
\begin{bchapter}
\bauthor{\bsnm{He}, \binits{K.}},
\bauthor{\bsnm{Zhang}, \binits{X.}},
\bauthor{\bsnm{Ren}, \binits{S.}},
\bauthor{\bsnm{Sun}, \binits{J.}}:
\bctitle{{Delving Deep into Rectifiers: Surpassing Human-Level Performance on
  ImageNet Classification}}.
In: \bbtitle{2015 {IEEE} International Conference on Computer Vision, {ICCV}
  2015, Santiago, Chile, December 7-13, 2015},
pp. \bfpage{1026}--\blpage{1034}
(\byear{2015}).
\doiurl{10.1109/ICCV.2015.123}
\end{bchapter}
\endbibitem

\bibitem[\protect\citeauthoryear{Paszke
  et~al.}{2019}]{DBLP:conf/nips/PaszkeGMLBCKLGA19}
\begin{bchapter}
\bauthor{\bsnm{Paszke}, \binits{A.}}, \betal:
\bctitle{{PyTorch: An Imperative Style, High-Performance Deep Learning
  Library}}.
In: \bbtitle{Advances in Neural Information Processing Systems 32: Annual
  Conference on Neural Information Processing Systems 2019, NeurIPS 2019,
  December 8-14, 2019, Vancouver, BC, Canada},
pp. \bfpage{8024}--\blpage{8035}
(\byear{2019})
\end{bchapter}
\endbibitem

\bibitem[\protect\citeauthoryear{Lecun et~al.}{1998}]{726791}
\begin{barticle}
\bauthor{\bsnm{Lecun}, \binits{Y.}},
\bauthor{\bsnm{Bottou}, \binits{L.}},
\bauthor{\bsnm{Bengio}, \binits{Y.}},
\bauthor{\bsnm{Haffner}, \binits{P.}}:
\batitle{{Gradient-based learning applied to document recognition}}.
\bjtitle{Proceedings of the IEEE}
\bvolume{86}(\bissue{11}),
\bfpage{2278}--\blpage{2324}
(\byear{1998})
\doiurl{10.1109/5.726791}
\end{barticle}
\endbibitem

\bibitem[\protect\citeauthoryear{Sandler et~al.}{2018}]{8578572}
\begin{bchapter}
\bauthor{\bsnm{Sandler}, \binits{M.}},
\bauthor{\bsnm{Howard}, \binits{A.}},
\bauthor{\bsnm{Zhu}, \binits{M.}},
\bauthor{\bsnm{Zhmoginov}, \binits{A.}},
\bauthor{\bsnm{Chen}, \binits{L.-C.}}:
\bctitle{{MobileNetV2: Inverted Residuals and Linear Bottlenecks}}.
In: \bbtitle{2018 IEEE/CVF Conference on Computer Vision and Pattern
  Recognition},
pp. \bfpage{4510}--\blpage{4520}
(\byear{2018}).
\doiurl{10.1109/CVPR.2018.00474}
\end{bchapter}
\endbibitem

\bibitem[\protect\citeauthoryear{Tan and Le}{2019}]{DBLP:conf/icml/TanL19}
\begin{bchapter}
\bauthor{\bsnm{Tan}, \binits{M.}},
\bauthor{\bsnm{Le}, \binits{Q.V.}}:
\bctitle{{EfficientNet: Rethinking Model Scaling for Convolutional Neural
  Networks}}.
In: \beditor{\bsnm{Chaudhuri}, \binits{K.}},
\beditor{\bsnm{Salakhutdinov}, \binits{R.}} (eds.)
\bbtitle{Proceedings of the 36th International Conference on Machine Learning,
  {ICML} 2019, 9-15 June 2019, Long Beach, California, {USA}}.
\bsertitle{Proceedings of Machine Learning Research},
vol. \bseriesno{97},
pp. \bfpage{6105}--\blpage{6114}
(\byear{2019})
\end{bchapter}
\endbibitem

\bibitem[\protect\citeauthoryear{Yi et~al.}{2024}]{YI2024101626}
\begin{barticle}
\bauthor{\bsnm{Yi}, \binits{D.-F.}},
\bauthor{\bsnm{Liu}, \binits{Q.}},
\bauthor{\bsnm{Ma}, \binits{R.-T.}},
\bauthor{\bsnm{Wang}, \binits{B.-L.}},
\bauthor{\bsnm{Liu}, \binits{H.-B.}},
\bauthor{\bsnm{Xie}, \binits{F.}},
\bauthor{\bsnm{Feng}, \binits{H.-B.}},
\bauthor{\bsnm{Feng}, \binits{Z.-K.}},
\bauthor{\bsnm{Zheng}, \binits{Y.-H.}}:
\batitle{{Star-XP: A simulation framework for Polar-2/ low energy X-ray
  polarization detector}}.
\bjtitle{SoftwareX}
\bvolume{25},
\bfpage{101626}
(\byear{2024})
\doiurl{10.1016/j.softx.2023.101626}
\end{barticle}
\endbibitem

\bibitem[\protect\citeauthoryear{Von~Mises}{1964}]{von1964mathematical}
\begin{bbook}
\bauthor{\bsnm{Von~Mises}, \binits{R.}}:
\bbtitle{Mathematical Theory of Probability and Statistics}.
\bpublisher{Academic press},
\blocation{New York and London}
(\byear{1964})
\end{bbook}
\endbibitem

\bibitem[\protect\citeauthoryear{Hundrieser et~al.}{2022}]{Hundrieser2022}
\begin{bbook}
\bauthor{\bsnm{Hundrieser}, \binits{S.}},
\bauthor{\bsnm{Klatt}, \binits{M.}},
\bauthor{\bsnm{Munk}, \binits{A.}}:
In: \beditor{\bsnm{SenGupta}, \binits{A.}},
\beditor{\bsnm{Arnold}, \binits{B.C.}} (eds.)
\bbtitle{{The Statistics of Circular Optimal Transport}},
pp. \bfpage{57}--\blpage{82}.
\bpublisher{Springer},
\blocation{Singapore}
(\byear{2022}).
\doiurl{10.1007/978-981-19-1044-9_4}
\end{bbook}
\endbibitem

\bibitem[\protect\citeauthoryear{Virtanen et~al.}{2020}]{2020SciPy-NMeth}
\begin{barticle}
\bauthor{\bsnm{Virtanen}, \binits{P.}}, \betal:
\batitle{{{SciPy} 1.0: Fundamental Algorithms for Scientific Computing in
  Python}}.
\bjtitle{Nature Methods}
\bvolume{17},
\bfpage{261}--\blpage{272}
(\byear{2020})
\doiurl{10.1038/s41592-019-0686-2}
\end{barticle}
\endbibitem

\bibitem[\protect\citeauthoryear{Flamary et~al.}{2021}]{flamary2021pot}
\begin{barticle}
\bauthor{\bsnm{Flamary}, \binits{R.}}, \betal:
\batitle{{POT: Python Optimal Transport}}.
\bjtitle{Journal of Machine Learning Research}
\bvolume{22}(\bissue{78}),
\bfpage{1}--\blpage{8}
(\byear{2021})
\end{barticle}
\endbibitem

\bibitem[\protect\citeauthoryear{Erdmann et~al.}{2023}]{Erdmann2023}
\begin{barticle}
\bauthor{\bsnm{Erdmann}, \binits{J.}},
\bauthor{\bsnm{Graaf}, \binits{A.}},
\bauthor{\bsnm{Mausolf}, \binits{F.}},
\bauthor{\bsnm{Nackenhorst}, \binits{O.}}:
\batitle{{SR-GAN for SR-gamma: super resolution of photon calorimeter images at
  collider experiments}}.
\bjtitle{The European Physical Journal C}
\bvolume{83}(\bissue{11}),
\bfpage{1001}
(\byear{2023})
\doiurl{10.1140/epjc/s10052-023-12178-3}
\end{barticle}
\endbibitem

\bibitem[\protect\citeauthoryear{Fu et~al.}{2024}]{Fu2024}
\begin{barticle}
\bauthor{\bsnm{Fu}, \binits{Z.}},
\bauthor{\bsnm{Grant}, \binits{C.}},
\bauthor{\bsnm{Krawiec}, \binits{D.M.}},
\bauthor{\bsnm{Li}, \binits{A.}},
\bauthor{\bsnm{Winslow}, \binits{L.A.}}:
\batitle{{Generative models for simulation of KamLAND-Zen}}.
\bjtitle{The European Physical Journal C}
\bvolume{84}(\bissue{6}),
\bfpage{651}
(\byear{2024})
\doiurl{10.1140/epjc/s10052-024-12980-7}
\end{barticle}
\endbibitem

\bibitem[\protect\citeauthoryear{He et~al.}{2016}]{7780459}
\begin{bchapter}
\bauthor{\bsnm{He}, \binits{K.}},
\bauthor{\bsnm{Zhang}, \binits{X.}},
\bauthor{\bsnm{Ren}, \binits{S.}},
\bauthor{\bsnm{Sun}, \binits{J.}}:
\bctitle{{Deep Residual Learning for Image Recognition}}.
In: \bbtitle{2016 IEEE Conference on Computer Vision and Pattern Recognition
  (CVPR)},
pp. \bfpage{770}--\blpage{778}
(\byear{2016}).
\doiurl{10.1109/CVPR.2016.90}
\end{bchapter}
\endbibitem

\bibitem[\protect\citeauthoryear{Dosovitskiy
  et~al.}{2021}]{DBLP:conf/iclr/DosovitskiyB0WZ21}
\begin{bchapter}
\bauthor{\bsnm{Dosovitskiy}, \binits{A.}}, \betal:
\bctitle{{An Image is Worth 16x16 Words: Transformers for Image Recognition at
  Scale}}.
In: \bbtitle{9th International Conference on Learning Representations, {ICLR}
  2021, Virtual Event, Austria, May 3-7, 2021}
(\byear{2021})
\end{bchapter}
\endbibitem

\end{thebibliography}

\end{document}